\pdfoutput=1
\documentclass[11pt,twoside,a4paper,cmspaper,final,collab]{cms-tdr}
\def\svnVersion{4181aff}\def\svnDate{2022/12/13}\def\cmsCernNoTag{CERN-EP-2021-206}\def\cmsCernDate{\today}\def\cmsMessage{Published in the European Physical Journal C as \href{http://dx.doi.org/10.1140/epjc/s10052-022-10897-7}{\doi{10.1140/epjc/s10052-022-10897-7}.}}
\begin{document}\cmsNoteHeader{SMP-18-013}

\providecommand{\cmsTable}[1]{\resizebox{\textwidth}{!}{#1}}
\ifthenelse{\boolean{cms@external}}{\providecommand{\cmsTableii}[1]{\resizebox{\columnwidth}{!}{#1}}}{\providecommand{\cmsTableii}[1]{#1}}
\newcommand{\pp}{\ensuremath{{\Pp\Pp}}}%
\newcommand{\rts}{\ensuremath{\sqrt{s}}}%
\newcommand{\IComb}    {I_{\text{comb}}}%

\newcommand{\MN}    {\ensuremath{\PGm\PGn}}%
\newcommand{\Wmn}   {\ensuremath{\PW \to \MN}}%
\newcommand{\EN}    {\Pe\PGn}%
\newcommand{\Wen}   {\ensuremath{\PW \to \EN}}%
\newcommand{\Wtn}   {\ensuremath{\PW \to \PGt\PGn}}%
\newcommand{\LN}    {\ensuremath{\ell\PGn}}%
\newcommand{\PWln}  {\ensuremath{\PW \to \LN}}%
\newcommand{\PWpln} {\ensuremath{\PWp \to \ell^+\PGn}}%
\newcommand{\PWmln} {\ensuremath{\PWm \to \ell^-\PAGn}}%

\newcommand{\Zee}   {\ensuremath{\PZ \to \EE}}%
\newcommand{\Zll}   {\ensuremath{\PZ \to \ell\ell}}%

\newcommand{\Vj}    {\ensuremath{\PV\hspace{-.25em}+\hspace{-.14em}\text{HF jets}}}
\newcommand{\Wj}    {\ensuremath{\PW\hspace{-.2em}+\hspace{-.14em}\text{jets}}}
\newcommand{\Zj}    {\ensuremath{\PZ\hspace{-.2em}+\hspace{-.14em}\text{jets}}}
\newcommand{\Wc}    {\ensuremath{\PW\hspace{-.25em}+\hspace{-.14em}\PQc}}
\newcommand{\Wb}    {\ensuremath{\PW + \PQb}}
\newcommand{\Wlight} {\ensuremath{\PW + {\PQu\PQd\PQs\Pg}}\xspace}
\newcommand{\Wcc}   {\ensuremath{\PW + \ccbar}}
\newcommand{\Wbb}   {\ensuremath{\PW + \bbbar}}
\newcommand{\WQQ}   {\ensuremath{\PW + {\PQQ\PAQQ} }\xspace}

\newcommand{\PWmc}  {\ensuremath{\PWm\hspace{-.25em}+\hspace{-.08em}\PQc}}
\newcommand{\PWpc}  {\ensuremath{\PWp\hspace{-.25em}+\hspace{-.08em}\PAQc}}
\newcommand{\SWpc}  {\sigma(\PWpc)}
\newcommand{\SWmc}  {\sigma(\PWmc)}
\newcommand{\SWc}   {\ensuremath{\sigma(\Wc)}}
\newcommand{\SWcdifflineeta} {\ensuremath{\rd\sigma(\Wc)/\rd\abs{\eta^\ell} }}
\newcommand{\SWcdifflinept}  {\ensuremath{\rd\sigma(\Wc)/\rd{\pt^\ell} }}
\newcommand{\Rcpm}           {\ensuremath{R_{\PQc}^{\pm}}}

\newcommand{\jet}   {\text{jet}}
\newcommand{\cjet}  {\PQc\text{ jet}\xspace}

\newcommand{\ppWc}   {\ensuremath{\Pp\Pp \to \PW\hspace{-.20em}+\hspace{-.14em}\PQc\hspace{-.14em}+\hspace{-.14em}\PX}}%
\newcommand{\ppWpc}  {\ensuremath{\Pp\Pp \to \PWp\hspace{-.24em}+\hspace{-.14em}\PAQc\hspace{-.14em}+\hspace{-.14em}\PX}}%
\newcommand{\ppWmc}  {\ensuremath{\Pp\Pp \to \PWm\hspace{-.24em}+\hspace{-.14em}\PQc\hspace{-.14em}+\hspace{-.14em}\PX}}%
\newcommand{\noppWc} {\ensuremath{\PW\hspace{-.25em}+\hspace{-.14em}\PQc}}%

\newcommand{\OSSS}   {\ensuremath{\text{OS-SS}}\xspace}%
\newcommand{\ptell}  {\pt^{\ell}}%
\newcommand{\etaell} {\eta^{\ell}}%

\newcommand{\ubar}{\PAQu}%
\newcommand{\dbar}{\PAQd}%
\newcommand{\sbar}{\PAQs}%

\newcommand{\Lambdacpm} {\ensuremath{\Lambda_{\PQc}^\pm}}
\DeclareRobustCommand{\PDstpmATLAS}{{\HepParticle{\PD}{}{(\ast)\pm}}\Xspace}

\newlength\cmsTabSkip\setlength{\cmsTabSkip}{1ex}

\newcommand{\convino} {{\textsc{Convino}}\xspace}
\newcommand{\PYTHIAsix} {{\textsc{pythia6}}\xspace}
\newcommand{\Hathor} {{\textsc{Hathor}}\xspace}

\newlength\cmsFigWidth
\ifthenelse{\boolean{cms@external}}{\setlength\cmsFigWidth{0.49\textwidth}}{\setlength\cmsFigWidth{0.65\textwidth}} 
\ifthenelse{\boolean{cms@external}}{\providecommand{\cmsLeft}{upper\xspace}}{\providecommand{\cmsLeft}{left\xspace}}
\ifthenelse{\boolean{cms@external}}{\providecommand{\cmsRight}{lower\xspace}}{\providecommand{\cmsRight}{right\xspace}}

\cmsNoteHeader{SMP-18-013} 
\title{Measurements of the associated production of a W boson and a charm quark in proton-proton collisions at \texorpdfstring{$\sqrt{s}=8\TeV$}{sqrt{s} = 8 TeV}}%
\titlerunning{W+c production in pp collisions at \texorpdfstring{$\sqrt{s}=8\TeV$}{sqrt{s} = 8 TeV}}

\date{\today}

\abstract{
Measurements of the associated production of a W boson and a charm ($\PQc$) quark in proton-proton collisions at a centre-of-mass energy of  8\TeV are reported. The analysis uses a data sample corresponding to a total integrated luminosity of 19.7\fbinv collected by the CMS detector at the LHC.  The W bosons are identified through their leptonic decays to an electron or a muon, and a neutrino. Charm quark jets are selected using distinctive signatures of charm hadron decays. The product of the cross section and branching fraction $\sigma(\Pp\Pp \to \PW + \PQc + \PX) \mathcal {B}(\PW \to \ell \Pgn)$, where $\ell = \Pe$ or $\PGm$,  and  the  cross  section  ratio $\sigma(\Pp\Pp \to {\PWp + \PAQc + \PX}) / \sigma(\Pp\Pp \to {\PWm + \PQc + \PX})$ are measured in a fiducial volume and differentially as functions of the pseudorapidity and of the transverse momentum of the lepton from the W boson decay. The results are compared with theoretical predictions. The impact of these measurements on the determination of the strange quark distribution is assessed. 
}

\hypersetup{
pdfauthor={CMS Collaboration},%
pdftitle={Measurements of the associated production of a W boson and a charm quark in proton-proton collisions at sqrt(s)= 8 TeV},%
pdfsubject={CMS},
pdfkeywords={CMS, SMP, W+charm}}

\maketitle 

\section{Introduction}

The CERN LHC has provided a large sample of proton-proton ($\pp$) collisions containing events with a vector boson (V) accompanied by one or more jets originating from heavy-flavour quarks
($\Vj$). Precise measurements of $\Vj$ observables can be used to test theoretical calculations of these processes 
and the modelling of $\Vj$ events in the currently available Monte Carlo (MC) event generator programs.

Measurements of $\Vj$ production also provide new input to the determination of the quark content of the proton. 
This information constrains the proton parton distribution functions (PDFs), a ubiquitous ingredient in many data analyses at LHC, 
and still an important source of systematic uncertainty (see \eg Ref.~\cite{tricoli2020vector} for a recent review). 
In this context, the measurements of the associated production of a {\PW} boson and a charm ($\PQc$) quark ($\Wc$ production) in proton-proton collisions at the LHC at $\rts=8\TeV$ 
presented in this paper provide new valuable information.

Measurements of $\Wc$ production in hadronic collisions at the\TeV scale were performed at the Tevatron by the CDF~\cite{CDF:2012mhm, CDF-VD} and 
D0~\cite{D0:2008ygk} Collaborations. 
The $\Wc$ process has been studied in $\pp$ collisions at the LHC at centre-of-mass energies of 7, 8 and 13\TeV by the 
CMS~\cite{CMS-PAPER-SMP-12-002, CMS-PAPER-SMP-17-014}, ATLAS~\cite{WplusCATLAS}, and LHCb~\cite{LHCb:2015bwt} experiments.

For the CMS measurement at $\sqrt{s}=7\TeV$ with integrated luminosity of about 5\fbinv, 
$\Wc$ candidates are identified through exclusive or semileptonic decays of charm hadrons inside a jet with transverse momentum of the 
jet larger than 25\GeV. 
The ATLAS analysis at the same centre-of-mass energy and similar integrated luminosity tags $\Wc$ events either 
by the presence of a muon from a semileptonic charm decay within a hadronic jet with transverse momentum larger than 25\GeV 
or by the reconstruction of a charm hadron exclusive decay with transverse momentum of the $\PDstpmATLAS$ candidate above 8\GeV. 
The CMS analysis at $\sqrt{s}=13\TeV$ with an integrated luminosity of 35.7\fbinv,  
uses the $\PDstp \to \PDz \PGpp$ with $\PDz \to \PKm \PGpp$ (plus the charge conjugated process) exclusive decay 
with transverse momentum of the $\PDstpm$ candidate above 5\GeV.
The LHCb measurement is based on integrated luminosities of 1 (2)\fbinv at $\sqrt{s}=7~(8) \TeV$, 
and uses tagging algorithms based on Boosted Decision Trees for the identification of $\PQc$ jets in conjunction with $\PQb$ jets.

We present in this paper the first measurement of the $\Wc$ production cross section at $\sqrt{s}=8\TeV$ in the central region.
The {\PW} boson is identified by a high transverse momentum isolated lepton ($\Pe, \PGm$) coming from its leptonic decay.
Fiducial cross sections are measured, both inclusively and differentially as functions 
of the absolute value of the pseudorapidity ($\etaell$) and, for the first time, the transverse momentum ($\ptell$) of the lepton from the {\PW} boson decay. 
Jets containing a $\PQc$ quark are identified in two ways: i) the identification of a muon inside the jet that comes from the semileptonic decay of a $\PQc$ flavoured hadron, 
and ii) a secondary vertex arising from a visible charm hadron decay.
The secondary-vertex $\cjet$ identification method, also newly introduced in this analysis, provides a large sample of $\Wc$ candidates. 
Measurements obtained in these four channels ({\Pe} and {\PGm} decay of {\PW} boson, $\cjet$ with muon or secondary vertex) are combined, 
resulting in reduced systematic uncertainties compared with previous CMS measurements.

\begin{figure*}[htbp]
  \centering
  \includegraphics[width=0.99\textwidth]{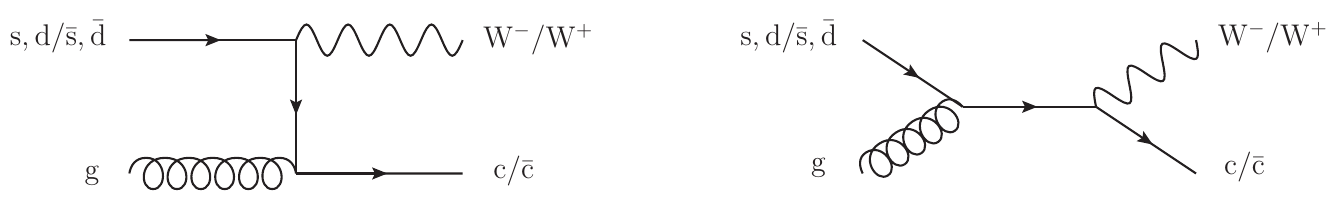}
  \caption{Leading order diagrams for the associated production of a {\PW} boson and a charm (anti)quark.}
  \label{fig:OSSS_diagram}
\end{figure*}

The study of $\Wc$ production at the LHC provides direct access 
to the strange quark content of the proton at the {\PW} boson mass energy scale~\cite{Baur}. 
The sensitivity comes from the dominance of the $\sbar\Pg \to \PWpc$ and $\PQs\Pg \to \PWmc$ contributions in the hard process, as depicted in Fig.~\ref{fig:OSSS_diagram}. 
The inclusion of strangeness-sensitive LHC measurements in global analyses of the proton PDFs has led to 
a significant reduction of the uncertainty in the strange quark PDF~\cite{Faura:2020oom}. 
The contribution of additional LHC $\Wc$ measurements will provide valuable input to further constrain the strange quark content of the proton.

A key property of $\Wc$ production is the opposite sign (OS) of the electric charges of the {\PW} boson and the $\PQc$ quark. 
Gluon splitting processes like $\cPq\cPaq^\prime \to \PW+\Pg \to$ $\PW+\PQc\PAQc$ also give rise to final states with 
an OS {\PW} boson and a $\PQc$ quark (antiquark), but with an additional $\PQc$ antiquark (quark) of the 
same sign (SS) electric charge as that of the {\PW} boson. 
In most of the background processes, it is equally probable to select events with OS electric charges as with SS,
whereas ${\cPq\Pg \to \PW+\PQc}$ only yields OS events.
Furthermore, distributions of the physical observables of OS and SS background events are expected to be the same, thus,  
the statistical subtraction of OS and SS distributions leads to an effective removal of these charge-symmetric backgrounds. 
This technique is referred to in the paper as \OSSS subtraction.
In the present analysis, the electric charges of the lepton from the {\PW} boson decay and the muon (or that assigned to the secondary vertex) inside the $\cjet$ 
are used to perform the \OSSS subtraction procedure.

{\tolerance=800 
The product of the cross sections and branching fraction $\sigma(\ppWpc)\mathcal{B}(\PWpln)$, 
$\sigma(\ppWmc)\mathcal{B}(\PWmln)$, their sum $\sigma(\ppWc)\mathcal{B}(\PWln)$, and the cross section ratio 
$\sigma(\ppWpc)/\sigma(\ppWmc)$, are measured at $\rts=8\TeV$.
They are abbreviated as $\SWpc$, $\SWmc$, $\SWc$, and $\Rcpm$.
The cross sections and cross section ratio are measured at the parton level in a fiducial region of phase space defined in terms of the kinematics of the lepton from the {\PW} boson 
($\ptell > 30\GeV$, and $\abs{\etaell} < 2.1$), 
and the $\PQc$ quark ($\pt^{\PQc} > 25\GeV$ and $\abs{\eta^{\PQc}} < 2.5$) with a separation between the $\PQc$ quark and the lepton 
$\Delta R({\PQc},\ell)= \sqrt{\smash[b]{(\Delta\eta)^2 +(\Delta\phi)^2}} > 0.5$.
The cross sections and cross section ratio are also measured differentially as functions of $\abs{\etaell}$ and $\ptell$.
\par}

The paper is structured as follows: the CMS detector is briefly described in Section~\ref{sec:CMS_det}, and the data and simulated samples used are
presented in Section~\ref{sec:samples}. Section~\ref{sec:event_sel} presents the selection of the signal sample.
Section~\ref{sec:syst_uncert} reviews the sources of systematic uncertainties and their impact on the measurements. 
The measurements of the fiducial $\Wc$ cross section and $\Rcpm$
are detailed in Section~\ref{sec:xsec_inc}, the differential measurements are reported in Section~\ref{sec:xsec_diff}, and a comparison with theoretical predictions is presented in Section~\ref{sec:theory}.
The details of the QCD analysis are described in Section~\ref{sec:qcd_analysis}.
Finally, the main results of the paper are summarized in Section~\ref{sec:summary}.

Tabulated results are provided in the HEPData record for this analysis~\cite{HepData}.

\section{The CMS detector~\label{sec:CMS_det}}

The central feature of the CMS apparatus is a superconducting solenoid
of 6\unit{m} internal diameter, providing a magnetic field of 3.8\unit{T}.
Within the solenoid volume are a silicon pixel
and strip tracker, a lead tungstate crystal electromagnetic calorimeter (ECAL),
and a brass and scintillator hadron calorimeter,
each composed of a barrel and two endcap sections.
Extensive forward calorimetry complements the coverage
provided by the barrel and endcap detectors.
The silicon tracker measures charged particles within the pseudorapidity
range $\abs{\eta}< 2.5$. It consists of 1440 silicon pixel
and 15\,148 silicon strip detector modules.
For particles of $1 < \pt < 10\GeV$ and $\abs{\eta} < 1.4$, the
track resolutions are typically 1.5\% in $\pt$ and 25--90 (45--150)\mum
in the transverse (longitudinal) impact parameter~\cite{CMS-PAPER-TRK-11-001}.
The electron momentum is estimated by combining the energy measurement
in the ECAL with the momentum measurement in the tracker. The momentum
resolution for electrons with $\pt \approx 45\GeV$ from $\Zee$ decays
ranges from 1.7\% for nonshowering electrons in the barrel region to 4.5\% for
showering electrons in the endcaps~\cite{CMS-PAPER-EGM-13-001}.
Muons are measured in the pseudorapidity range $\abs{\eta}< 2.4$,
using three technologies: drift tubes,
cathode strip chambers, and resistive plate chambers.
Matching muons to tracks measured in the silicon tracker
results in a relative transverse momentum resolution for muons
with $20<\pt<100\GeV$ of 1.3--2.0\% in the barrel and
better than 6\% in the endcaps.
The \pt resolution in the barrel is better than 10\% for muons with \pt up to 1\TeV~\cite{CMS-PAPER-MUO-10-004}.
For muons with $1<\pt<25\GeV$, the relative transverse momentum resolution is 1.2--1.7\% in the barrel and 2.5--4.0\%
in the endcaps~\cite{CMS-PAPER-TRK-11-001}.
Events of interest are selected using a two-tiered trigger system~\cite{Khachatryan:2212926}. 
The first level, composed of custom hardware processors, uses information from the calorimeters and muon detectors to select events at a rate 
of around 100\unit{kHz} within a fixed latency of about 4\mus. The second level, known as the high-level trigger, consists of a farm of processors running 
a version of the full event reconstruction software optimized for fast processing, and reduces the event rate to around 1\unit{kHz} before data storage. 
A more detailed description of the CMS detector, together with a definition of the coordinate system used and the basic kinematic variables, can be found in Ref.~\cite{Chatrchyan:2008zzk}.

\section{Data and simulated samples~\label{sec:samples}}

The data were collected by the CMS experiment during 2012 in $\pp$ collisions at a centre-of-mass energy of 8\TeV
with an integrated luminosity of 19.7\fbinv.

{\tolerance 800 
Samples of simulated events are produced with MC event generators, both for the signal process and for the main
backgrounds. They are normalized to the integrated luminosity of the data sample using their respective cross sections.
A sample of $\Wj$ events is generated with \MADGRAPH v5.1.3.30~\cite{Alwall:2011uj},
interfaced with \PYTHIA v6.4.26~\cite{Pythia6} for parton showering and hadronization using the MLM~\cite{Alwall:2007fs, Alwall:2008qv} jet matching scheme.
The \MADGRAPH generator produces parton-level events with a vector boson and up to four partons on the basis of a leading order (LO) matrix-element calculation.
The generator uses the parton distribution function (PDF) set CTEQ6L~\cite{Pumplin:2002vw}, which is reweighted to the next-to-next-to-leading-order (NNLO) PDF set MSTW2008NNLO~\cite{Martin:2009iq}.
A sample of $\Zj$ events, which includes the exchange of a virtual photon, is generated with \MADGRAPH interfaced with \PYTHIAsix with the same conditions as for the $\Wj$ event sample.
They are normalized to the inclusive $\PW$ and $\PZ$ production cross sections evaluated at NNLO with \FEWZ3.1~\cite{Li:2012wna},
using the MSTW2008NNLO PDF set.
\par}

{\tolerance 800 
Background samples of top (t) quark events ($\ttbar$ and single top) are generated at next-to-leading-order (NLO) with \POWHEG v1.0~\cite{Campbell:2014kua, Nason:2004rx, Frixione:2007vw, Alioli:2010xd}, interfaced with \PYTHIAsix and using the CT10~\cite{Gao:2013xoa} PDF set. 
The $\ttbar$ cross section is taken at NNLO from Ref.~\cite{Czakon:2013goa}. The t-channel single-top cross section is calculated at NLO with \Hathor v2.1~\cite{Aliev:2010zk,Kant:2014oha}
and the $\PQt\PW$ and s-channel cross sections are taken at NNLO from Ref.~\cite{Kidonakis:2013zqa}. 
Diboson (VV) production ($\PW\PW$, $\PW\PZ$, and $\PZ\PZ$ processes) is modelled with samples of events
generated with \PYTHIAsix and the CTEQ6L1 PDF set.
Their cross sections are evaluated at NLO with \MCFM 6.6~\cite{Campbell:2010ff}, using the MSTW2008NLO PDF set.
For all simulations, the \PYTHIAsix parameters for the underlying event modelling are set to the Z2$^{\ast}$ tune~\cite{Chatrchyan:2013gfi, Khachatryan:2015pea}. Final state QED radiation is modelled by \PYTHIAsix. 
\par}

{\tolerance 800 
Simulated events are weighted to correct the charm quark fragmentation fractions into the weakly decaying hadrons \PDpm, \PDz/\PADz, \PDpms and 
$\Lambdacpm$ in \PYTHIAsix, to match the combination of measurements given in Ref.~\cite{Lisovyi:2015uqa}.
An additional event weight correcting the decay branching fractions larger than 1\% of \PDz{}/\PADz and \PDpm mesons is introduced to make them agree with more recent values~\cite{Pythia82,newPDG}. 
These decay modes altogether represent about 70\% of the total \PDz{}/\PADz and \PDpm decay rate. 
The remaining \PDz{}/\PADz and \PDpm decay modes are globally adjusted to keep the normalization of the decay branching fractions to unity. 
The \PDz{}/\PADz and \PDpm mesons constitute about 80\% of the total number of produced charm hadrons, 
thus approximately 56\% of the charm sample is  corrected by this adjustment. 
\par}

Generated events are processed through a \GEANTfour-based~\cite{Agostinelli:2002hh} CMS detector simulation and trigger emulation.
Simulated events are then reconstructed using the same algorithms used to reconstruct collision data.

The simulated samples incorporate additional $\pp$ interactions in the same bunch crossing (pileup) to reproduce the experimental conditions.
Simulated events are weighted so that the pileup distribution
matches the measured one, with an average of about 21 $\pp$ interactions per bunch crossing.

The simulated trigger, reconstruction, and selection efficiencies are corrected to match those observed in the data.
Lepton efficiencies ($\epsilon_{\ell}$) are evaluated with data samples of dilepton events in the {\PZ} boson mass peak with the ``tag-and-probe'' method~\cite{CMS-PAPER-EWK-10-005}, and
correction factors $\epsilon_{\ell}^\text{data}/\epsilon_{\ell}^{\mathrm\mathrm{MC}}$, binned in $\pt$ and $\eta$
of the leptons, are computed. These corrections are typically close to 1\% for muons and 3\% for electrons, with no relevant dependence 
on the $\pt$ and $\eta$ of the lepton. 

{\tolerance 800 
The simulated signal sample is composed of {\PW} bosons accompanied by jets originating from $\PQb$, $\PQc$, and light quarks (or antiquarks) and gluons.
Simulated $\Wj$ events are classified according to the flavour of the generated partons. 
A $\Wj$ event is categorized as $\Wc$ if a single charm quark with $\pt>15\GeV$ is generated in the hard process. 
Otherwise, it is classified as $\Wb$ if at least one $\PQb$ quark with $\pt>15\GeV$ is generated. 
Remaining events are labelled as $\Wcc$ if at least a $\PQc\PAQc$ quark-antiquark pair is present in the event, or as $\Wlight$ if no $\PQc$ or $\PQb$ quarks are produced.  
The contribution from the $\Wcc$ process is expected to vanish after \OSSS subtraction.
\par}

\section{Event reconstruction and selection~\label{sec:event_sel}}

Jets, missing transverse momentum, and related quantities are determined using the CMS particle-flow (PF) reconstruction algorithm~\cite{Sirunyan:2017ulk},
which aims to reconstruct and identify each individual particle in an event, with an optimized combination of information from the various elements of the 
CMS detector. 

Jets are built from PF candidates using the anti-\kt clustering algorithm~\cite{Cacciari:2008gp, Cacciari:2011ma} with a distance parameter $R = 0.5$.
The energy and momentum of the jets are corrected, as a function of the jet $\pt$ and $\eta$, to account for the nonlinear response of the
calorimeters and for the presence of pileup interactions~\cite{CMS-PAPER-JME-10-011, CMS-PAPER-JME-13-004}.
Jet energy corrections are derived using samples of simulated events and further adjusted using dijet, photon+jet, and {\PZ}+jet events in data.

Electron and muon candidates are reconstructed following standard CMS procedures~\cite{CMS-PAPER-EGM-13-001, CMS-PAPER-MUO-10-004}.
The missing transverse momentum vector $\ptvecmiss$ is the projection of the negative vector sum of the momenta, onto the plane perpendicular to the beams,  
of all the PF candidates. 
The $\ptvecmiss$ is modified to include corrections to the energy scale of the reconstructed jets in the event.
The missing transverse momentum, $\ptmiss$, is defined as the magnitude of the $\ptvecmiss$
vector, and it is a measure of the transverse momentum of particles leaving the detector undetected~\cite{CMS-PAPER-JME-12-002}.

The primary vertex of the event, representing the hard interaction, is selected among the reconstructed vertices as the one with the highest sum of the transverse momenta squared of the tracks associated with it.

\subsection{Selection of \texorpdfstring{$\PW$}{W} boson events~\label{sec:Wsel}}

Events with a high-$\pt$ lepton from the {\PW} boson decay are selected online by a trigger algorithm that requires the presence of an electron with $\pt > 27\GeV$ or a 
muon with $\pt>24\GeV$.
The analysis follows the selection criteria used in Ref.~\cite{CMS-PAPER-SMP-14-020} and requires the presence of a
high-$\pt$ isolated lepton in the pseudorapidity region $\abs{\eta} < 2.1$. The $\pt$ of the lepton must exceed 30\GeV. 

The combined isolation $\IComb$ is used to quantify the additional hadronic activity around the selected leptons.
It is defined as the sum of the transverse momentum of neutral hadrons, photons and the $\pt$ of charged hadrons in a cone with
$\Delta R = \sqrt{\smash[b]{(\Delta\eta)^2 +(\Delta\phi)^2}}<0.3$ $(0.4)$ around the electron (muon) candidate, excluding the contribution from the lepton itself.
Only charged particles originating from the primary vertex are considered in the sum to minimize the contribution from pileup interactions.
The contribution of neutral particles from pileup vertices is estimated and subtracted from $\IComb$.
For electrons, this contribution is evaluated with the jet area method described in Ref.~\cite{jet_area}; for muons,
it is taken to be half the sum of the $\pt$ of all charged particles in the cone originating from pileup vertices.
The factor one half accounts for the expected ratio of neutral to charged particle production in hadronic interactions.
The electron (muon) candidate is considered to be isolated when $\IComb/\ptell < 0.15$ $(0.12)$.
Events with a second isolated lepton with $\ptell>20\GeV$ and $\abs{\eta} < 2.1$, and opposite charge to the lepton from the {\PW} candidate 
are discarded to reduce the contribution from $\Zj$ and $\ttbar$ events.

The transverse mass ($\mT$) of the lepton and $\ptvecmiss$ is defined as,
\begin{linenomath*}
\begin{equation*}
  \mT \equiv \sqrt{{2~\ptell~\ptmiss~[1-\cos(\phi_\ell-\phi_{\ptmiss})]}},
\end{equation*}
\end{linenomath*}
where $\phi_\ell$ and $\phi_{\ptmiss}$ are the azimuthal angles of the lepton momentum and the $\ptvecmiss$ vector, respectively. 
Events with $\mT < 55\GeV$ are discarded from the analysis to suppress the contamination from QCD multijet events. 
The remaining contribution after \OSSS subtraction is negligible.

\subsection{Selection of \texorpdfstring{$\Wc$}{W + charm} events~\label{sec:Wjetssel}}

A $\Wj$ sample is selected from the sample of {\PW} boson events by additionally requiring the presence of at least one jet with transverse momentum ($\pt^{\jet}$) larger than 25\GeV
in the pseudorapidity region $\abs{\eta^{\jet}}<2.5$.
Jets are not selected if they have a separation $\Delta R ({\text{jet}},\ell)$ $<0.5$ in the $\eta$-$\phi$ space between the jet axis and the selected isolated lepton.

Hadrons with $\PQc$ quark content decay weakly with lifetimes of the order of $10^{-12}\unit{s}$ and mean decay lengths larger than 100\unit{$\mu$m}
at the LHC energies. Secondary vertices well separated from the primary vertex are reconstructed from the tracks of their charged decay products.
In a sizeable fraction of the decays (${\approx}$10--15\%~\cite{newPDG}) there is a muon in the final state.
We make use of these properties and focus on the following two signatures to identify jets originating from a $\PQc$ quark:
{\tolerance 800 
\begin{itemize}
\item[$\centerdot$] \textbf{ {Semileptonic (SL) channel}}, a well-identified muon inside the jet coming from the semileptonic decay of a charm hadron.
\item[$\centerdot$] \textbf{ {Secondary vertex (SV) channel}}, a reconstructed displaced secondary vertex inside the jet.
\end{itemize}
When an event fulfils the selection requirements of both topologies, it is assigned to the SL channel, which has a higher purity. 
Thus, the SL and the SV categories are mutually exclusive, \ie, the samples selected in each channel are statistically independent. 
The event selection process is summarized in Table~\ref{table:Summary_cuts} for the four analysis categories, the $\PW$ boson decay channels to electron or muon,
and the SL and SV charm identification channels.
\par}

These two signatures are also features of weakly decaying $\PQb$ hadrons. Events from physical processes producing $\PQb$ jets accompanied by a {\PW} boson
will be abundantly selected in the two categories. The most important source of background events is $\ttbar$ production, 
where a pair of {\PW} bosons and two $\PQb$ jets are produced in the decay of the top quark-antiquark pair. 
This final state mimics the analysis topology when at least one of the {\PW} bosons decays leptonically, and there is an identified muon or
a reconstructed secondary vertex inside one of the $\PQb$ jets. 
However, this background is effectively suppressed by the \OSSS subtraction. 
The chance to identify a muon or a secondary vertex inside the $\PQb$ jet with opposite or same charge than the charge 
of the {\PW} candidate is identical, thus delivering an equal number of OS and SS events. 

Top quark-antiquark events where one of the {\PW} bosons decays hadronically
into a $\PQc\PAQs$ (or $\PAQc\PQs$) quark-antiquark pair may result in additional event candidates if the SL or SV signature originates from the $\PQc$ jet. 
This topology produces real OS events, which contribute to an additional background after \OSSS subtraction. 
Similarly, single top quark production also produces real OS events, but at a lower level because of the smaller production cross section. 

The production of a {\PW} boson and a single $\PQb$ quark through the process $\Pq\Pg \to \Wb$, similar to the one sketched in Fig.~\ref{fig:OSSS_diagram},  
produces actual OS events, but it is heavily Cabibbo-suppressed and its contribution to the analysis is negligible. The other source of a {\PW} boson and a $\PQb$ quark is $\Wbb$ events 
where the $\bbbar$ pair originates from gluon splitting and only one of the two $\PQb$ jets is identified. These events are also charge symmetric as it is equally likely to
identify the $\PQb$ jet with the same or opposite charge than that of the {\PW} boson and its contribution cancels out after the \OSSS subtraction.

\begin{table*}[htbp]
 \centering
  \topcaption{Summary of the selection requirements for the four analysis categories.}
  \begin{tabular}{llr}
   \multicolumn {3}{c}{ $\PW$ + jets selection }  \\
   \hline
   Channel & $\Wen$ & $\Wmn$ \\
   \hline
   Lepton $\pt^{\ell}$ & \multicolumn{2}{c}{$> 30\GeV$} \\
   Lepton $\abs{\eta^{\ell}}$ & \multicolumn{2}{c}{$< 2.1$} \\
   Lepton isolation $\IComb/\pt^{\ell}$ & $< 0.15$ & $< 0.12$  \\
   Transverse mass $\mT$ & \multicolumn{2}{c}{$> 55\GeV$} \\
   Jet $\pt^{\jet}$ & \multicolumn{2}{c}{$> 25\GeV$} \\
   Jet $\abs{\eta^{jet}}$ & \multicolumn{2}{c}{$< 2.5$} \\
   $\Delta R ({\text{jet}},\ell)$ & \multicolumn{2}{c}{$> 0.5$} \\ [\cmsTabSkip]
   \multicolumn {3}{c}{ $\Wc$ -- SL channel }  \\
   \hline
   Muon in jet $\pt^{\PGm}$ & \multicolumn{2}{c}{$< 25\GeV$} \\
   Muon in jet $\abs{\eta^{\PGm}}$ & \multicolumn{2}{c}{$< 2.1$} \\
   Muon in jet $\pt^{\PGm}/\pt^{\jet}$ & \multicolumn{2}{c}{$< 0.6$} \\
   Muon in jet isolation $\IComb/\pt^{\ell}$ & \multicolumn{2}{c}{$> 0.2$} \\
   Muon in jet IPS & \NA & $>1$ \\
   \multirow{2}{*}{Muon in jet $m_{\mu\mu}$} & \multirow{2}{*}{\NA} & $>12\GeV$ \& \\
                                             &      & $\notin[70,110\GeV]$ \\ [\cmsTabSkip]
   \multicolumn {3}{c}{ $\Wc$ -- SV channel }  \\
   \hline
   Secondary-vertex displacement significance, SV 3D & \multicolumn{2}{c}{$> 3.5$} \\
   Corrected secondary-vertex mass, $m_\text{SV}^\text{corr}$ & \multicolumn{2}{c}{$> 0.55\GeV$} \\
   Secondary-vertex charge & \multicolumn{2}{c}{$\neq 0$} \\
   \hline
  \end{tabular}
  \label{table:Summary_cuts}
\end{table*}

\subsubsection{Event selection in the SL channel~\label{sec:Wsel_SL}}

The $\Wc$ events with a semileptonic charm hadron decay are identified by a reconstructed muon among the constituents of any of the selected jets.
The muon candidate has to satisfy the same reconstruction and identification quality criteria as those imposed on the muons from the {\PW} boson decay, has to be reconstructed in the region 
$\abs{\eta} < 2.1$ with $\pt^{\PGm}<25\GeV$ and $\pt^{\PGm}/\pt^{\jet}<0.6$,
and it must not be isolated from hadron activity, $\IComb/\pt^{\PGm}>0.2$.
No minimum $\pt$ threshold is explicitly required, but the muon reconstruction algorithm sets a natural threshold around 3\GeV (2\GeV) in the barrel (endcap) region, 
since the muon must traverse the material in front of the muon detector and travel deep enough into the muon 
system to be reconstructed and satisfy the identification criteria.
If more than one such muon is identified, the one with the highest $\pt$ is selected.
The electric charges of the muon in the jet and the lepton from the {\PW} boson decay determine whether the event is treated as OS or SS.
Semileptonic decays into electrons are not selected because of the high background in identifying electrons inside jets.

Additional requirements are applied for the event selection in the $\Wmn$ channel, because the selected sample is affected by a sizeable contamination from dimuon $\Zj$ events.
Events with a dimuon invariant mass close to the {\PZ} boson mass peak ($70<m_{\PGm\PGm}<110\GeV$) are discarded.
Furthermore, the invariant mass of the muon pair must be larger than 12\GeV to suppress the background from low-mass resonances.

Finally, if the muon in the jet candidate comes from a semileptonic decay of a charm hadron, its associated track
is expected to have a significant impact parameter, defined as the projection in the transverse plane of the vector between the primary vertex and the muon trajectory at its
point of closest approach.
To further reduce the $\Zj$ contamination in the  $\Wmn$ channel, we require the impact parameter significance (IPS) of the muon in the jet, 
defined as the muon impact parameter divided by its uncertainty, to be larger than 1. 

The above procedure results in an event yield of $52\,179 \pm 451$ ($32\,071 \pm 315$), after \OSSS subtraction, in the $\Wen$ ($\Wmn$) channel  where the quoted uncertainty is statistical. 
The smaller yield in the  $\Wmn$ channel is mainly due to the requirement on the IPS of the muon inside the jet, 
which is solely applied to this channel.
Table~\ref{fractiontable0} shows the flavour composition of the selected sample according to simulation. The fraction of $\Wc$ signal events is around 80\%. 
The dominant background arises from $\ttbar$ production (around 8\%), where one of the {\PW} bosons produced in the decay of the top quark
pair decays leptonically and the other hadronically with a $\PQc$ quark in the final state. 
The contribution from $\ttbar$ events where one of the top quarks is out of the acceptance of the detector is estimated with the simulated sample to be negligible. 
Figure~\ref{fig:muon_in_jet_pt} shows the distributions after \OSSS subtraction of the IPS (left) and $\pt$ (right) of the muon inside the jet for events in the selected sample. 
The difference between data and simulation in the high-$\pt$ region in Fig.~\ref{fig:muon_in_jet_pt}, right ($\pt\gtrsim 20\GeV$), 
is related to a similar behaviour observed in the $\pt^{\PGm}$/$\pt^{\jet}$ distribution. 
Differences are significantly reduced by reweighting the simulation with weights extracted from the $\pt^{\PGm}$/$\pt^{\jet}$ 
distribution to make the corresponding simulation description match the data.
\begin{linenomath*}
 \begin{figure*}[htb]
  \centering
   \includegraphics[width=0.48\textwidth]{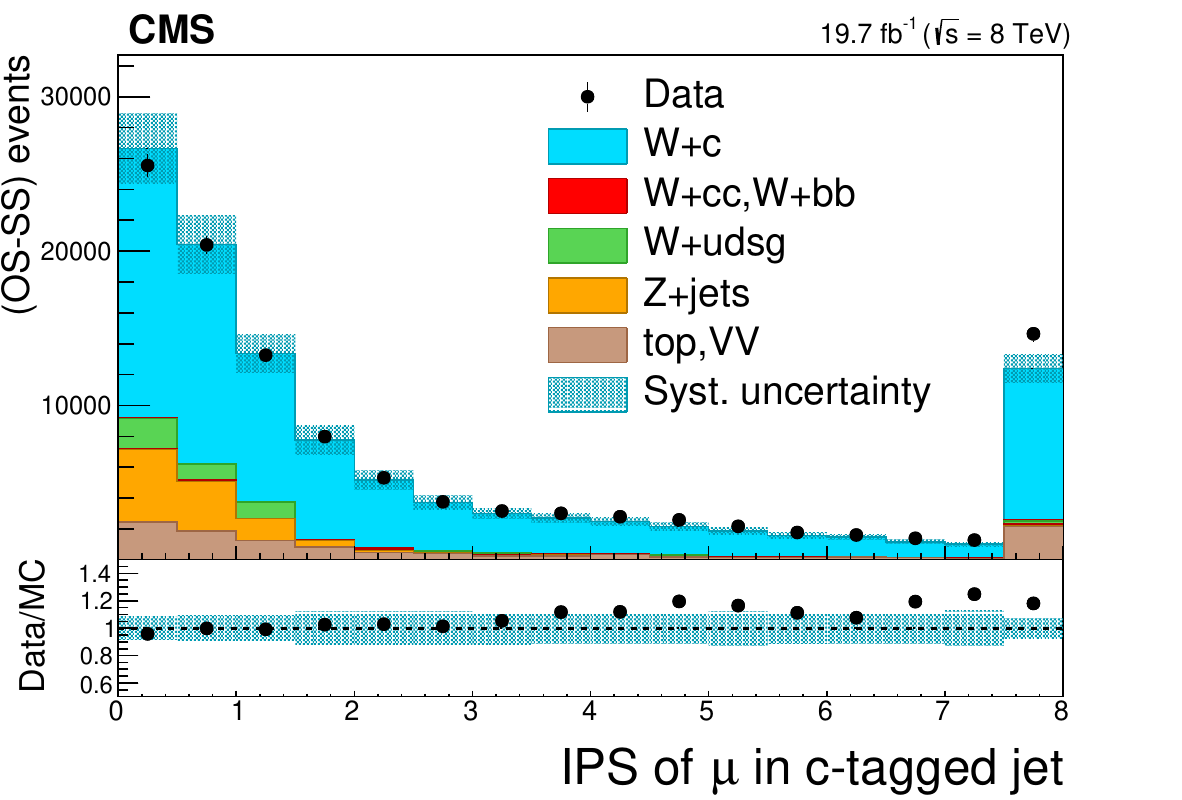}
   \includegraphics[width=0.48\textwidth]{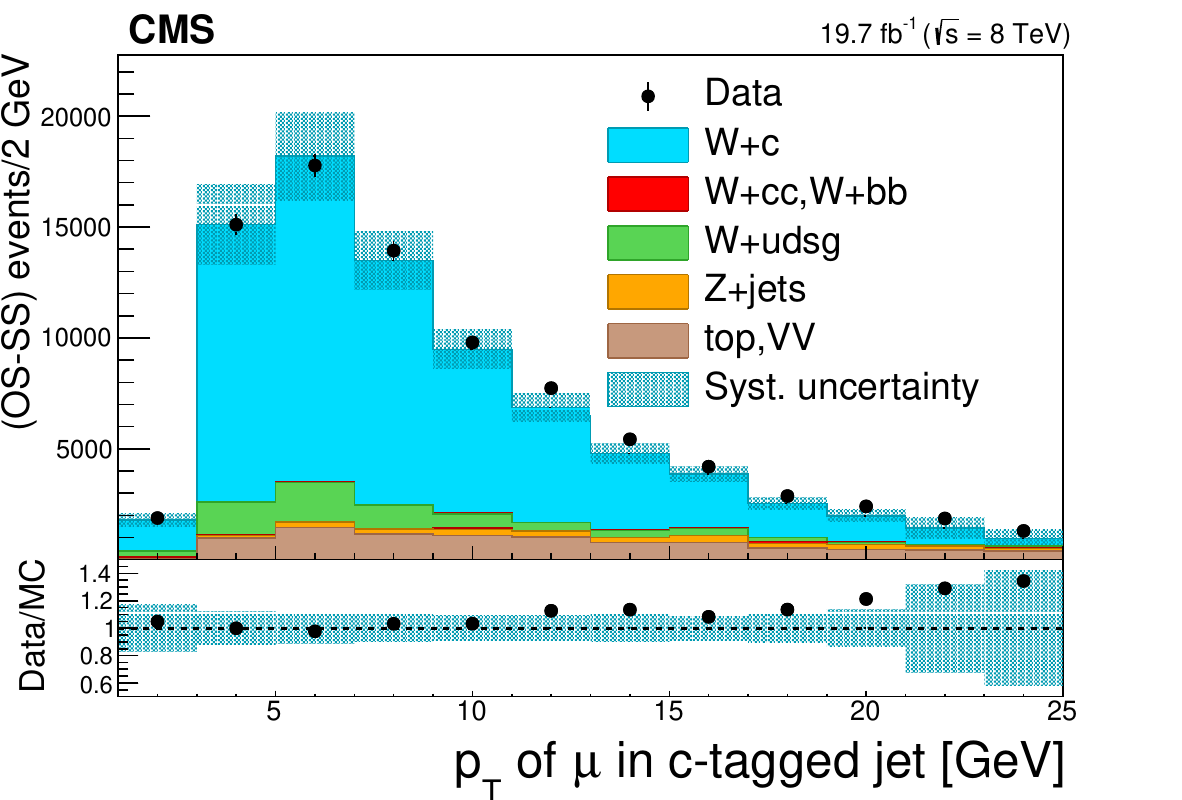}
   \caption{Distributions after \OSSS subtraction of the impact parameter significance, IPS, (left) and $\pt$ (right),  
   of the muon inside the $\cjet$ for events in the SL sample, summing up the contributions of the two {\PW} boson decay channels. 
   The IPS distribution is shown after all selection requirements except the one on this variable. 
   The last bin of the distribution includes all events with $\text{IPS}>7.5$. 
   The $\pt$ distribution includes the selection requirement $\text{IPS}>1.0$ for the $\Wmn$ channel.
   The contributions of the various processes are estimated with the simulated samples. 
   Vertical bars on data points represent statistical uncertainty in the data. 
   The hatched areas represent the sum in quadrature of statistical and systematic uncertainties in the MC simulation.
   The ratio of data to simulation is shown in the lower panels. 
   The uncertainty band in the ratio includes the statistical uncertainty in the data, and the statistical and systematic uncertainties in the MC simulation.}
   \label{fig:muon_in_jet_pt}
\end{figure*}
\end{linenomath*}
\begin{linenomath*}
 \begin{table*}[htbp]
 \centering
  \topcaption{Simulated flavour composition (in \%) of the SL sample after the selection summarized in Table~\ref{table:Summary_cuts} and \OSSS subtraction, 
  for the electron and muon decay channels of the {\PW} boson. {\WQQ} is the sum of the contributions of {\Wcc} and {\Wbb}; 
  its negative value is an effect of the OS-SS subtraction. Quoted uncertainties are statistical only.}
  \renewcommand{\arraystretch}{1.2}
  \begin{tabular}{lcccccccc}
    SL channel      & \Wc  & \WQQ & \Wlight & \Zj  & \ttbar & single t  & $\PV\PV$  \\
    \hline
    $\Wen$ & $84.1 \pm 0.9$ &  $-0.6 \pm 0.4$ & $4.5 \pm 0.7$ & $0.5 \pm 0.2$ & $8.3 \pm 0.4$ & $2.3 \pm 0.1$ & $0.9 \pm 0.1$ \\
    $\Wmn$ & $78.7 \pm 1.1$ & $\,\,\,\,\, 0.1 \pm 0.5$ & $3.1 \pm 0.7$ & $7.0 \pm 0.2$ & $7.7 \pm 0.5$ & $2.5 \pm 0.1$ & $0.9 \pm 0.1$ \\ 
  \end{tabular}
  \label{fractiontable0} 
\end{table*}
\end{linenomath*}

\subsubsection{Event selection in the SV channel~\label{sec:Wsel_SV}}

An independent $\Wc$ sample is selected looking for secondary decay vertices of charm hadrons within the reconstructed jets. 
Displaced secondary vertices are reconstructed with either the simple secondary vertex (SSV)~\cite{CMS-PAPER-BTV-12-001} or the
inclusive vertex finder (IVF)~\cite{Khachatryan:2011wq, Chatrchyan:2013zja} algorithms.
Both algorithms follow the adaptive vertex fitter technique~\cite{Adaptive_Vertex} to  construct a secondary vertex, but differ in the tracks used. 
The SSV algorithm takes as input the tracks constituting the jet; the IVF algorithm starts from a displaced track with respect to the primary vertex
(\textit{seed} track) and tries to build a vertex from nearby tracks in terms of their separation distance in three dimensions and their angular separation
around the seed track. IVF vertices are then associated with the closest jet in a cone of $\Delta R=0.3$. 
Tracks used for the reconstruction of both secondary vertices must have $\pt>1\GeV$ to avoid misreconstructed or poorly reconstructed tracks.
 
If there are several jets with a secondary vertex, only the jet with the highest transverse momentum is selected. 
If more than one secondary vertex within a jet is reconstructed, the one with the highest transverse momentum, computed from its associated tracks, is considered.  

To ensure that the secondary vertex is well separated from the primary one, we require the secondary-vertex displacement significance,
defined as the three dimensional (3D) distance between the primary and the secondary vertices, divided by its uncertainty, to be larger than 3.5.

{\tolerance 800 
We define the corrected secondary-vertex mass, $m_\text{SV}^\text{corr}$, as
the invariant mass of all charged particles associated with the secondary vertex, assumed to be pions, $m_\text{SV}$, 
corrected for additional particles, either charged or neutral, that may have been produced but were not reconstructed~\cite{Aaij:2015yqa}: 
\begin{linenomath*}
\begin{equation*}
 m_\text{SV}^\text{corr} = \sqrt{m^2_\text{SV} + p^2_\text{SV} \sin^2 \theta}  + p_\text{SV} \sin \theta,
\end{equation*}
\end{linenomath*}
where $p_\text{SV}$ is the modulus of the vectorial sum of the momenta of all charged particles associated with the secondary vertex, 
and $\theta$ is the angle between the momentum vector sum and the vector from the primary to the secondary vertex.
The corrected secondary-vertex mass is thus, the minimum mass the long-lived hadron can have that is consistent with the direction of flight.
To reduce the contamination of jets not produced by the hadronization of a heavy-flavour quark (light-flavour jet background), $m_\text{SV}^\text{corr}$ must be larger than 0.55\GeV.
\par}

Vertices reconstructed with the IVF algorithm are considered first. 
If no IVF vertex is selected, SSV vertices are searched for, thus providing additional event candidates. 

For charged charm hadrons, the sum of the charges of the decay products reflects the charge of the $\PQc$ quark. 
For neutral charm hadrons, the charge of the closest hadron produced in the fragmentation process can indicate the charge of the $\PQc$ quark~\cite{chargedeterminationreference2, chargedeterminationreference}. 
Hence, to classify the event as OS or SS, we scrutinize the charge of the secondary vertex and of the nearby tracks. 
We consider the SV as positively (negatively) charged if the sum of the charges of the constituent tracks is larger (smaller) than zero.
If the secondary vertex charge is zero, we take the charge of the primary vertex track closest to the direction of the secondary vertex (given by the sum of the momentum of the constituent tracks). 
We only consider primary vertex tracks with $\pt>0.3\GeV$ and within an angular separation, $\Delta R < 0.1$, from the secondary vertex direction.
If non zero charge cannot be assigned, the event is rejected.

{\tolerance 800 
In about 45\% of the selected events, the reconstructed charge of the secondary vertex is zero, and in 60\% of them, a charge can be assigned from the primary vertex track. 
According to the simulation, the charge assignment is correct in 70\% of the cases, both for charged and neutral secondary vertices. 
\par}

After \OSSS subtraction, we obtain an event yield of  $118\,625 \pm 947$ ($132\,117 \pm 941$) in the $\Wen$ ($\Wmn$) channel. 
Table~\ref{fractiontable} shows the flavour composition of the selected sample, as predicted by the simulation. The purity of the $\Wc$ signal events is about 75\%. 
The dominant background comes from $\Wlight$ jets (around 15\%), mostly from the processes
${\PQu \Pg \to \PWp + \PQd }$  and ${\PQd \Pg \to \PWm + \PQu}$, which are OS. Figure~\ref{fig:sv_in_jet_pt} shows the distributions after \OSSS subtraction of the secondary vertex displacement significance 
and the corrected secondary-vertex mass for data and simulation. 
\begin{linenomath*}
 \begin{table*}[htbp]
  \centering
  \topcaption{Simulated flavour composition (in \%) of the SV sample after the selection summarized in Table~\ref{table:Summary_cuts}, including OS-SS subtraction, 
   for the electron and muon {\PW} boson decay channels. {\WQQ} is the sum of the contributions of {\Wcc} and {\Wbb}. Quoted uncertainties are statistical only.}
  \renewcommand{\arraystretch}{1.2}
  \begin{tabular}{lcccccccc}
    SV channel      & \Wc  & \WQQ & \Wlight & \Zj & \ttbar & single t  & $\PV\PV$  \\
    \hline
    $\Wen$ & $74.9 \pm 1.1$ & $0.4 \pm 0.4$ & $15.1 \pm 0.9$ & $1.8 \pm 0.2$ & $3.5 \pm 0.3$ & $3.2 \pm 0.1$ & $1.1 \pm 0.1$ \\
    $\Wmn$ & $75.1 \pm 1.0$ & $0.4 \pm 0.4$ & $16.0 \pm 0.9$ & $0.7 \pm 0.2$ & $3.3 \pm 0.3$ & $3.5 \pm 0.1$ & $1.0 \pm 0.1$ \\
  \end{tabular}
  \label{fractiontable} 
 \end{table*}
\end{linenomath*}
\begin{linenomath*}
 \begin{figure*}[!tb]
  \centering
  \includegraphics[width=0.48\textwidth]{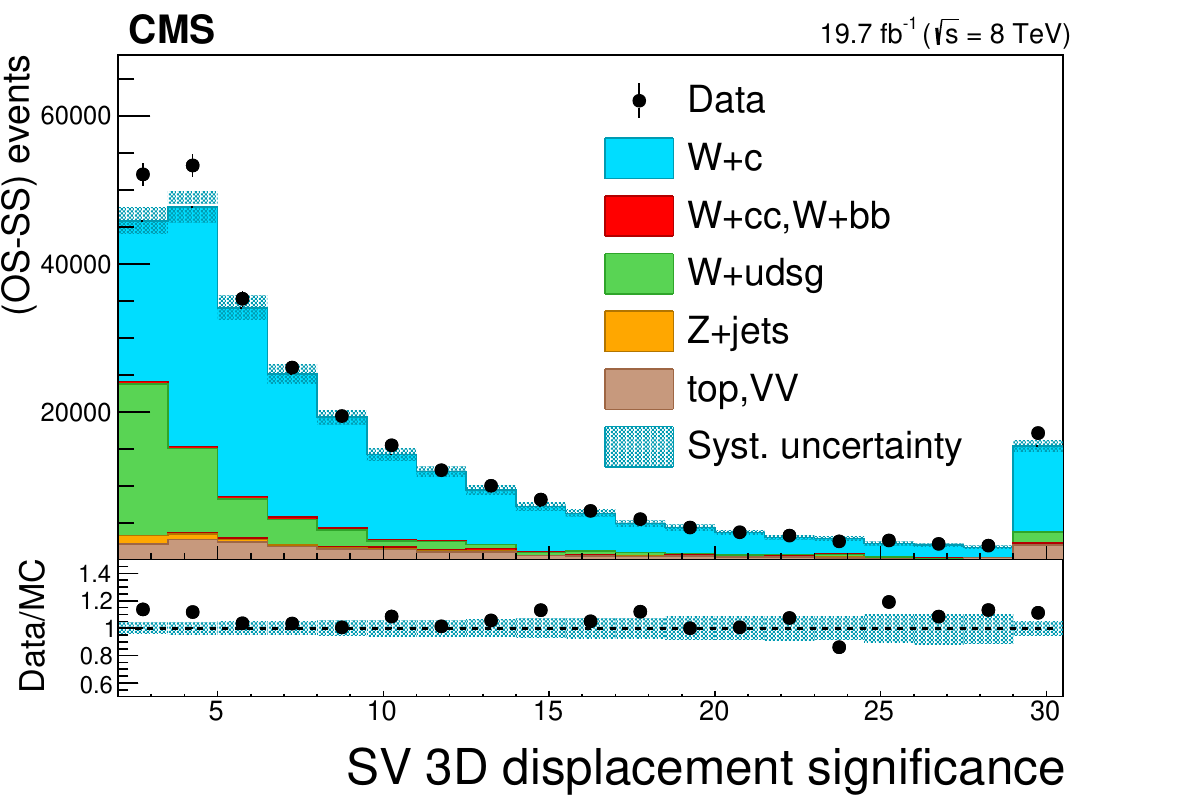}
  \includegraphics[width=0.48\textwidth]{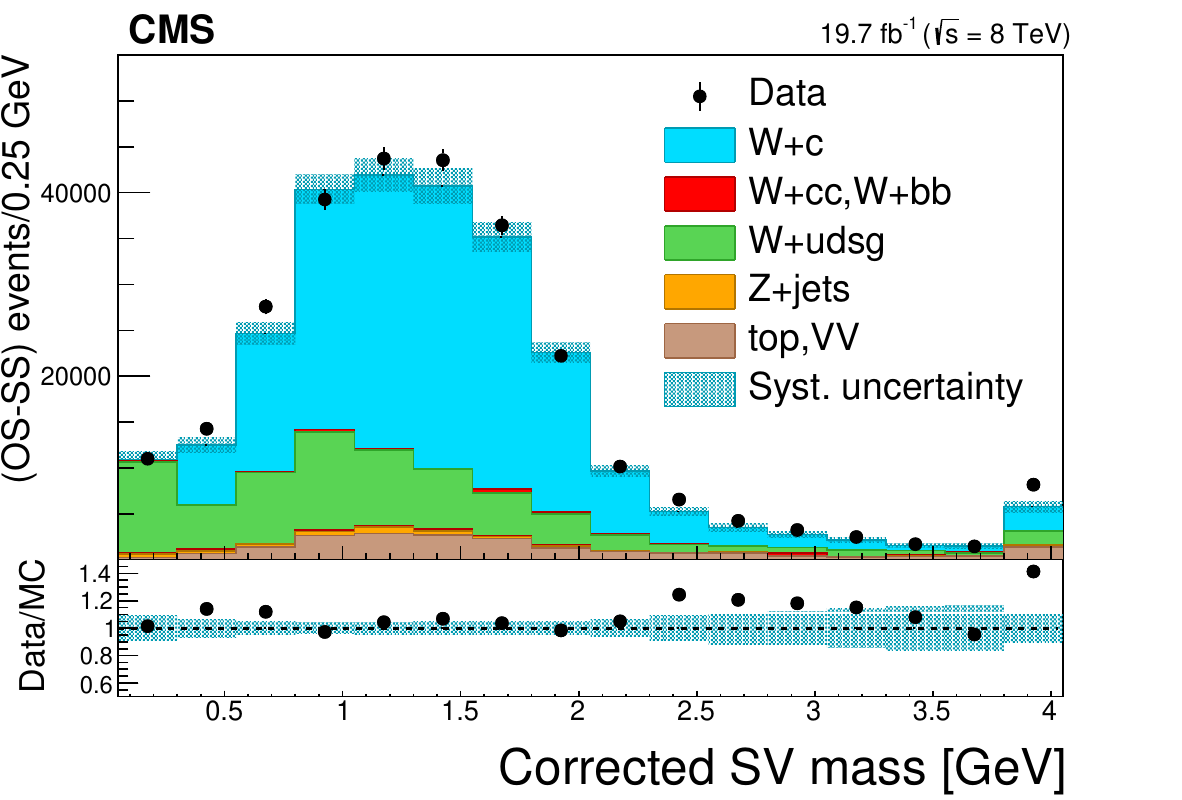}
  \caption{Distributions after \OSSS subtraction of the secondary-vertex displacement significance (left) and
  corrected secondary-vertex mass (right). 
  For each distribution all selection requirements are applied except the one on the displayed variable. 
  The last bin of each plot includes all events beyond the bin.
  The contributions from all processes are estimated with the simulated samples.
  Vertical bars on data points represent the statistical uncertainty in the data. 
  The hatched areas represent the sum in quadrature of statistical and systematic uncertainties in the MC simulation.
  The ratio of data to simulation is shown in the lower panels. 
  The uncertainty band in the ratio includes the statistical uncertainty in the data, and the statistical and systematic uncertainties in the MC simulation.}
  \label{fig:sv_in_jet_pt}
 \end{figure*}
\end{linenomath*}

The distributions from the MC simulations are corrected for known discrepancies between data and simulation in the secondary vertex reconstruction.
The events of the SL sample are used to compute data-to-simulation scale factors for the efficiency of charm identification through the reconstruction of 
a SV~\cite{EPJC78, CMS-PAPER-BTV-16-002}.
The fraction of events in the SL sample with a secondary vertex is computed for data and simulation, and the ratio of
data to simulation is applied as a scale factor to simulated $\Wc$ signal events in the SV sample.
The scale factor is $0.94 \pm 0.03$, where the uncertainty includes the statistical and systematic effects.
The systematic uncertainty includes contributions from the uncertainties in the pileup description, jet energy scale and resolution, 
lepton efficiencies, background subtraction, and modelling of charm production and decay fractions in the simulation.
The dependence of the scale factor on the $\pt$ of the jet is included when computing differential cross sections, as explained in Section~\ref{sec:xsec_diff}.

A jet $\pt$- and $\eta$-dependent correction factor between 1.0 and 1.2 is applied to the $\Wlight$ component of the $\Wj$ simulation to account for 
inaccuracies in the description of light-flavour jet contamination entering the signal. 
Those values correspond to data/simulation correction factors for light jets being misidentified as heavy-flavour jets, as computed in Ref.~\cite{CMS-PAS-BTV-13-001}.

\section{Systematic uncertainties~\label{sec:syst_uncert}}

The impact of various sources of uncertainty in the measurements is estimated by recalculating the cross sections and cross section ratio 
with the relevant parameters varied up and down by one standard deviation of their uncertainties. 
Most sources of systematic uncertainty equally affect $\SWpc$ and $\SWmc$ measurements, thus, their effects largely cancel in the cross section ratio.
We discuss first the uncertainties in the determination of the fiducial cross section in the four channels. 
The uncertainties in the cross section ratio are summarized at the end of the section.
The most relevant sources of systematic uncertainties in the differential cross sections are further discussed in Section~\ref{sec:xsec_diff}.

The combined uncertainty in the lepton trigger, reconstruction, and identification efficiencies results in a cross section uncertainty of 1.3 and 0.8\% 
for the $\Wen$ and $\Wmn$ channel, respectively. 
The uncertainty in the efficiency of the identification of muons inside jets is approximately 3\%,
according to dedicated studies in multijet events~\cite{CMS-PAPER-MUO-10-004}, which directly translates into an equivalent uncertainty in the measured cross section in the SL channels.

The probability of lepton charge misassignment is studied with data using $\Zll$ events reconstructed with same- or opposite-sign leptons. 
The charge misidentification probability for muons is negligible ($<10^{-4}$). For the electrons, it is ${\approx}$0.4\%, which propagates 
into a negligible uncertainty in the cross section measurements. 

The effects of the uncertainty in the jet energy scale and the jet energy resolution are assessed by varying the corresponding 
correction factors within their uncertainties, according to the results of dedicated CMS studies~\cite{CMS-PAPER-JME-10-011, CMS-PAPER-JME-13-004}. 
The resulting uncertainty is below 1.5\%. 
The uncertainty from a $\ptvecmiss$ mismeasurement in the event is estimated by
smearing the simulated $\ptvecmiss$ distribution to match that in data. The resulting uncertainty in the cross section is less than 0.2\%.
Uncertainties in the pileup modelling are calculated using a modified pileup profile obtained by changing the mean number of interactions by $\pm5\%$.
This variation covers the uncertainty in the $\pp$ inelastic cross section and in the modelling of the pileup simulation. 
It results in less than 1\% uncertainty in the cross section measurements.   

The measured average of the inclusive charm quark semileptonic branching fractions is ${\mathcal{B}}(\PQc\to\ell) = 0.096\pm 0.004$~\cite{newPDG},
while the exclusive sum of the individual contributions from all weakly decaying charm hadrons is $0.086\pm 0.004$~\cite{Lisovyi:2015uqa,newPDG}. 
The average of these two values, ${\mathcal{B}}(\PQc\to\ell) = 0.091 \pm 0.003$, is consistent with the \PYTHIA value used in our simulations (9.3\%).
We assign a 5\% uncertainty in the SL channel to cover both central values within one standard deviation.
For the SV channel, remaining inaccuracies in the charm hadron branching fractions in the \PYTHIAsix simulation are covered by a systematic uncertainty (2.6\%) 
equal to the change in the cross section caused by the correction of \PDz/\PADz and \PDpm decay branching fractions, as described in Section~\ref{sec:samples}. 
The systematic effect of the uncertainty in the charm quark fragmentation fractions is set to be equal to the change in the cross section (1.2\%) caused by the correction 
procedure described in Section~\ref{sec:samples}. This uncertainty is assigned to both the SL and SV channels.

To account for inaccuracies in the simulation of the energy fraction of the charm quark carried by the charm hadron in the fragmentation process, 
we associate a systematic uncertainty computed by weighting the simulation to match the distribution of an experimental observable representative of that quantity.
We use the distribution of the muon transverse momentum divided by the jet transverse momentum, $\pt^{\PGm}$/$\pt^{\jet}$, for the SL channel, and 
the secondary vertex transverse momentum divided by the jet transverse momentum, $\pt^{\text{SV}}$/$\pt^{\jet}$, for the SV channel. 
This procedure results in an uncertainty in the cross section of ${\approx}1\%$ in the SL channel and ${\lesssim}0.5\%$ in the SV channel.

The uncertainty in the scale factor correcting the SV reconstruction efficiency in simulation propagates into a systematic uncertainty of 2.2\% in the cross section.

The modelling of the simulation of the secondary vertex charge assignment efficiency is studied with data using the subset of the events of the SL sample
where a displaced secondary vertex has also been identified.  The requirement of a reconstructed secondary vertex in the SL sample increases the 
$\Wc$ signal contribution to 95\%. 
The charge of the secondary vertex is tested against the charge of the muon inside the jet, which is taken as a reference. 
The uncertainty in the SV charge determination is estimated as the difference in the rate obtained in data and simulation of correct SV charge assignment 
and results in a 1.2\% uncertainty in the cross section.

{\tolerance 800 
The uncertainty in the determination of the background processes is thoroughly evaluated. The \OSSS subtraction procedure efficiently suppresses the 
contribution from background processes that produce equal amounts of OS and SS candidates, thus rendering the measurements largely insensitive to the modelling of these backgrounds.
This is the case of $\ttbar$ production with the subsequent leptonic decay of the two {\PW} bosons, which is completely removed. 
We have checked with data how efficiently the \OSSS subtraction procedure eliminates these charge symmetric $\ttbar$ events.
A $\ttbar$-enriched control sample is selected by requiring a pair of high-$\pt$ isolated leptons of different flavour, $\Pe$-$\PGm$, with 
opposite charge, following the same lepton selection criteria as in the $\Wc$ analysis. Events with at most two reconstructed jets with $\pt>30\GeV$ are selected.
A nonisolated muon or a secondary vertex inside one of the jets is required. 
The charge of the highest-$\pt$ isolated lepton and the charge of the muon in the jet or the secondary vertex are compared to classify the event as OS or SS. 
The test is repeated taking separately the highest-$\pt$ lepton of the two possible lepton flavours and charges.
A reduction down to less than 1\% is observed in all cases after \OSSS subtraction.
This behaviour is well reproduced in the simulation.
\par}

Some background contribution is expected from $\ttbar$ events where one of the {\PW} bosons decays leptonically, 
and the other one decays hadronically into a $\PQc\PAQs$ ($\PAQc\PQs$) pair. These are genuine OS events. 
The accuracy of the simulation to evaluate this contribution is checked with data using a semileptonic $\ttbar$-enriched sample selected by requiring a high-$\pt$ isolated lepton ($\Pe$ or $\PGm$) 
fulfilling the criteria of the $\Wc$ selection, and at least four jets in the event, one of them satisfying either the SL or SV selection. 
The relative charge of the muon in the jet or the secondary vertex with respect to the lepton from the {\PW} decay determines the event to be OS or SS. 
The number of events after \OSSS subtraction in the simulation and in data agree better than 10\%.
This difference is assigned as the uncertainty in the description of the semileptonic $\ttbar$ background. The effect on the fiducial $\Wc$ cross section is smaller than 1\%. 

The uncertainty in the contribution from single top quark processes is estimated by varying the normalization of the samples according to the uncertainties in the
theoretical cross sections, $\sim$5--6\%. It produces a negligible effect on the measurements.

The contribution from $\Zj$ events is only relevant in the $\Wmn$ channel of the SL category, amounting to ${\sim}7\%$ of the selected events.
The level of agreement between data and the $\Zj$ simulation is studied in the region of the {\PZ} boson mass peak, $70<m_{\PGm\PGm}<110\GeV$, 
which is excluded in the signal analysis, applying the same selection procedure as for the signal sample, except for the invariant mass requirement; 
a difference of about 15\% is observed.
This discrepancy is assigned as a systematic uncertainty, assuming the same mismodelling outside the {\PZ} mass peak region. The effect on the cross section is about 1\%.

An additional systematic uncertainty is assigned to account for a possible mismodelling of the $\Wlight$ background.
The systematic uncertainty is evaluated by using simulation correction factors, as presented in Section~\ref{sec:Wsel_SV}, associated with different misidentification probabilities. The uncertainty in the $\Wlight$ contribution is ${\approx}10\%$, which translates into a 1\% uncertainty in the cross section. 

The \OSSS subtraction removes almost completely the contribution from gluon splitting processes to the selected sample. 
We have estimated that a possible mismodelling up to three times the 
experimental uncertainty in the gluon splitting rate into $\ccbar$ quark pairs~\cite{Alephgccbar,Alephgbbbar} 
has a negligible impact on the measurements.

The signal sample is generated with \MADGRAPH and \PYTHIAsix using the CTEQ6L1 PDF and weighted to NNLO PDF set MSTW2008NNLO.
The effect from the PDF uncertainty is estimated using other NNLO PDF sets (CT10 and NNPDF2.3~\cite{Ball:2012cx}). 
The resulting uncertainty in the cross section is small (${\lesssim}1\%$).
Following the prescription of the individual PDF groups, the PDF uncertainty is of the same order.

In the signal modelling, no uncertainties are included in the simulation of higher-order terms in perturbative QCD (parton shower) 
or nonperturbative effects (hadronization, underlying event). 
The $\OSSS$ subtraction technique removes the contributions to $\Wc$ production coming from charm quark-antiquark pair production, 
rendering the measurement insensitive to those effects. 

The statistical uncertainty in the determination of the selection efficiency using the simulated samples is 2\% for the SL channel and 1\% for the SV channel, 
and is propagated as an additional systematic uncertainty.
The uncertainty in the integrated luminosity is 2.6\%~\cite{CMS-PAS-LUM-13-001}.

The total systematic uncertainty in the $\Wc$ cross section is 7\% for the measurements in the SL channels, and 5\% for those in the SV channels.

Most of the systematic uncertainties cancel out in the measurement of the cross section ratio $\Rcpm$. 
This is the case of uncertainties related to lepton reconstruction and identification efficiencies, 
secondary vertex reconstruction, charm hadron fragmentation and decay fractions, 
and integrated luminosity determination. All other sources of uncertainty have a limited effect. The most relevant source of systematic uncertainty 
is the statistical uncertainty in the determination with the simulation of the selection efficiencies separately for the samples of $\PWp$ and $\PWm$ bosons.  
The total systematic uncertainty in the measurement of $\Rcpm$ in the SL channels is 3.5\%, and 2.5\% in the SV channels.

\section{Fiducial \texorpdfstring{$\Wc$}{W + c} cross section and \texorpdfstring{$(\PWpc)/(\PWmc)$}{(W+ + cbar)/(W- + c)} cross section ratio~\label{sec:xsec_inc}}

Cross sections are unfolded to the parton level using the $\Wc$ signal reference as defined in the \MADGRAPH generator at the hard-scattering level. 
Processes where a charm-anticharm quark pair is produced in the hard interaction are removed from the signal definition. 
To minimize acceptance corrections, the measurements are restricted to a phase space
that is close to the experimental fiducial volume with optimized sensitivity for the investigated processes:
a lepton with $\ptell>30\GeV$ and $\abs{\etaell} < 2.1$,
together with a $\PQc$ quark with $\pt^{\PQc} > 25\GeV$ and $\abs{\eta^{\PQc}} < 2.5$.
The $\PQc$ quark parton should be separated from the lepton of the {\PW} boson candidate
by a distance $\Delta R({\PQc},\ell)>0.5$. 

The measurement of the $\Wc$ cross section is performed independently in four different channels: the two charm identification SL and SV channels, and using {\PW} boson decay to electrons or muons.
For all channels under study, the $\Wc$ cross section is determined using the following expression:
\begin{linenomath*}
\begin{equation}
\SWc = \frac{Y_{\text{sel}}(1-f_{\text{bkg}})}{\mathcal{C} \, \mathcal{L}},
\label{eq:W_c_data}
\end{equation}
\end{linenomath*}
where $Y_{\text{sel}}$ is the selected event yield in data and $f_{\text{bkg}}$ the fraction of remaining background events, 
both after the selection process summarized in Table~\ref{table:Summary_cuts}, and \OSSS subtraction. 
The fraction $f_{\text{bkg}}$ is estimated from simulation.
The signal yield, $Y_{\text{sel}}(1-f_{\text{bkg}})$, is presented in Table~\ref{table:Cross_sectionsall}.  

The factor $\mathcal{C}$ corrects for losses in the selection process of $\Wc$ events produced in the fiducial region at parton level. 
It also subtracts the contributions from events outside the measurement fiducial region and from $\Wc$ events with $\Wtn$, $\PGt\to\Pe + \PX$ or $\PGt\to\PGm + \PX$. 
It is calculated, using the sample of simulated signal events, as the ratio between the event yield of the selected $\Wc$ sample (according to the procedure 
described in Sections~\ref{sec:Wsel_SL} and~\ref{sec:Wsel_SV} and after \OSSS subtraction) 
and the number of $\Wc$ events satisfying the phase space definition at parton level.
The values of the $\mathcal{C}$ factors are also given in Table~\ref{table:Cross_sectionsall}. The uncertainties quoted in the table include
statistical and the associated systematic effects as discussed in Section~\ref{sec:syst_uncert}.
The different values of $\mathcal{C}$ reflect the different reconstruction and selection efficiencies in the four channels.
In the SL channel, only about 3\% of the signal charm hadrons generated in the fiducial region of the analysis produce a muon in their decay 
with enough momentum to reach the muon detector and get reconstructed. 
In the SV channel, only about 6\% of the events with a charm hadron decay remain after SV reconstruction, SV charge assignment and \OSSS subtraction. 
The remaining inefficiency, accounted for in the $\mathcal{C}$ correction factors, is due to selection criteria of the samples.
According to the simulation, the contribution to the cross section of events with $\mT<55\GeV$ is around 20\%. 
No uncertainty is assigned to the modelling of this extrapolation.  
The integrated luminosity of the data is denoted by $\mathcal{L}$.

Finally, the fiducial $\Wc$ production cross section 
computed with Eq.~(\ref{eq:W_c_data}) in the SL and SV channels for the electron and muon decay channels separately is shown in the last column of Table~\ref{table:Cross_sectionsall}. 
Statistical and systematic uncertainties are quoted.
\begin{table*}[htbp] 
 \centering 
  \topcaption{Results in the SL (upper) and SV (lower) channels for the $\Wen$ and $\Wmn$ decays separately. 
           Here $Y_{\text{sel}}(1-f_{\text{bkg}})$ is the estimate for the signal event yield after background subtraction, 
           $\mathcal{C}$ is the acceptance times efficiency correction factor, and $\SWc$ is the measured production cross section.} 
  \renewcommand{\arraystretch}{1.2}
  \begin{tabular}{cccc} 
   \multicolumn {4}{c}{ SL channel }  \\ 
   \hline 
   Channel & $Y_{\text{sel}}(1-f_{\text{bkg}})$ &  $\mathcal{C}$ [\%] & $\SWc$ [pb] \\ 
   \hline 
   $\Wen$ & $43\,873 \pm 379$ & $1.95 \pm 0.03 \stat \pm 0.11 \syst$ & $113.3 \pm 1.2 \stat \pm 8.2 \syst$ \\ 
   $\Wmn$ & $25\,252 \pm 248$ & $1.11 \pm 0.03 \stat \pm 0.06 \syst$ & $115.7 \pm 1.4 \stat \pm 8.7 \syst$ \\ [\cmsTabSkip] 
   \multicolumn {4}{c}{ SV channel }  \\ 
   \hline 
   Channel & $Y_{\text{sel}}(1-f_{\text{bkg}})$ &  $\mathcal{C}$ [\%] & $\SWc$ [pb] \\ 
   \hline  
   $\Wen$ & $88\,899 \pm 710$ & $3.75 \pm 0.05 \stat \pm 0.15 \syst$ & $120.2 \pm 1.3 \stat \pm 6.4 \syst$ \\ 
   $\Wmn$ & $99\,167 \pm 706$ & $4.29 \pm 0.05 \stat \pm 0.17 \syst$ & $117.3 \pm 1.1 \stat \pm 6.2 \syst$ \\ 
  \end{tabular} 
  \label{table:Cross_sectionsall} 
\end{table*} 

The $\PWpc$ and $\PWmc$ cross sections are also measured independently using Eq.~(\ref{eq:W_c_data}) after splitting 
the sample according to the charge of the lepton from the {\PW} boson decay, and the cross section ratio is computed. 
The corresponding numbers are summarized in Table~\ref{table:Cross_sectionspm}. 
The overall yield of $\PWmc$ is expected to be slightly larger than that of $\PWpc$ due to the small contribution, at a few percent level, 
of $\Wc$ production from the Cabibbo-suppressed processes $\dbar\Pg \to \PWpc$ and $\PQd\Pg \to \PWmc$; 
this contribution is not symmetric because of the presence of down valence quarks in the proton. 
\begin{table*}[htbp] 
 \centering 
  \topcaption{Measured production cross sections $\SWpc$, $\SWmc$, and their ratio, $\Rcpm$, in the SL (upper) and SV (lower) channels for the electron and muon {\PW} boson decay modes.} 
  \ifthenelse{\boolean{cms@external}}{
  \renewcommand{\arraystretch}{1.2}
  \begin{tabular}{cccc} 
   \multicolumn {4}{c}{ SL channel }  \\ 
   \hline 
   Channel & $\SWpc$ [pb] & $\SWmc$ [pb] & $\Rcpm$\\ 
   \hline 
   $\Wen$ & $55.9 \pm 0.9\stat \pm 4.1\syst$ & $57.3 \pm 0.8\stat \pm 4.3$\syst & $0.976 \pm 0.020\stat \pm 0.034$\syst \\ 
   $\Wmn$ & $56.4 \pm 1.1\stat \pm 4.2\syst$ & $58.7 \pm 1.0\stat \pm 4.6$\syst & $0.961 \pm 0.024\stat \pm 0.036$\syst \\ [\cmsTabSkip] 
   \multicolumn {4}{c}{ SV channel }  \\ 
   \hline 
   Channel & $\SWpc$ [pb] & $\SWmc$ [pb] & $\Rcpm$\\ 
   \hline 
   $\Wen$ & $59.2 \pm 0.9\stat \pm 3.3\syst$ & $61.0 \pm 0.9\stat \pm 3.4\syst$ & $0.970 \pm 0.021\stat \pm 0.025\syst$  \\ 
   $\Wmn$ & $58.3 \pm 0.8\stat \pm 3.2\syst$ & $57.7 \pm 0.8\stat \pm 3.1\syst$ & $1.010 \pm 0.019\stat \pm 0.025\syst$ \\ 
  \end{tabular} 
   }{
  \cmsTable{
  \renewcommand{\arraystretch}{1.2}
  \begin{tabular}{cccc} 
   \multicolumn {4}{c}{ SL channel }  \\ 
   \hline 
   Channel & $\SWpc$ [pb] & $\SWmc$ [pb] & $\Rcpm$\\ 
   \hline 
   $\Wen$ & $55.9 \pm 0.9\stat \pm 4.1\syst$ & $57.3 \pm 0.8\stat \pm 4.3$\syst & $0.976 \pm 0.020\stat \pm 0.034$\syst \\ 
   $\Wmn$ & $56.4 \pm 1.1\stat \pm 4.2\syst$ & $58.7 \pm 1.0\stat \pm 4.6$\syst & $0.961 \pm 0.024\stat \pm 0.036$\syst \\ [\cmsTabSkip] 
   \multicolumn {4}{c}{ SV channel }  \\ 
   \hline 
   Channel & $\SWpc$ [pb] & $\SWmc$ [pb] & $\Rcpm$\\ 
   \hline 
   $\Wen$ & $59.2 \pm 0.9\stat \pm 3.3\syst$ & $61.0 \pm 0.9\stat \pm 3.4\syst$ & $0.970 \pm 0.021\stat \pm 0.025\syst$  \\ 
   $\Wmn$ & $58.3 \pm 0.8\stat \pm 3.2\syst$ & $57.7 \pm 0.8\stat \pm 3.1\syst$ & $1.010 \pm 0.019\stat \pm 0.025\syst$ \\ 
  \end{tabular} 
   }
   }
  \label{table:Cross_sectionspm} 
\end{table*}

Results obtained for the $\Wc$ cross sections and cross section ratios in the different channels are consistent within uncertainties, and 
are combined to improve the precision of the measurement. The \convino~\cite{Convino} tool, which is used to perform the combination,
is a maximum-likelihood approach including correlations between uncertainties within and between measurements.
Systematic uncertainties arising from a common source and affecting several measurements are considered as fully correlated.
In particular, all systematic uncertainties are assumed fully correlated between the electron and muon channels, except those related to the lepton reconstruction. 
The combined cross section and cross section ratio are:
\begin{linenomath*}
  \begin{equation*}
  \begin{aligned}
      \sigma(\noppWc) & = 117.4 \pm 0.6\stat \pm 5.6\syst \unit{pb}, \\
      \Rcpm           & = 0.983 \pm 0.010\stat \pm 0.017\syst. 
  \end{aligned}
  \end{equation*}
\end{linenomath*}

The contribution of the various sources of systematic uncertainty to the combined cross section is shown in Table~\ref{tabSFcsys}.
For each of the sources in the table, the quoted uncertainty is computed as the difference in quadrature between the uncertainty of the 
nominal combination and the one of a combination with that uncertainty fixed to the value returned by \convino.
\begin{table}[htbp]
  \centering 
  \topcaption{Impact of the sources of systematic uncertainty in the combined $\SWc$ measurement.}
  \renewcommand{\arraystretch}{1.2}
  \begin{tabular}{ l  c } 
   Source & Uncertainty [\%] \\
   \hline
   Lepton efficiency & 0.7\\
   Jet energy scale and resolution & 0.8 \\
   $\ptmiss$ resolution & 0.3 \\
   Pileup modelling & 0.4 \\
   $\PGm$ in jet reconstruction efficiency & 0.9 \\
   Secondary vertex reconstruction efficiency & 1.8 \\
   Secondary vertex charge determination & 1.0 \\
   Charm fragmentation and decay fractions & 2.6 \\
   Charm fragmentation functions & 0.3 \\
   Background subtraction & 0.8 \\
   PDF & 1.0 \\
   Limited size of MC samples & 0.6 \\
   Integrated luminosity & 2.6 
  \end{tabular} 
  \label{tabSFcsys}
\end{table} 

A prediction of the $\Wc$ cross section is obtained with the \MADGRAPH simulation sample.
It is estimated by applying the phase space definition requirements to the generator-level quantities:
a lepton from the {\PW} boson decay with $\ptell>30\GeV$ and $\abs{\etaell} < 2.1$;
a generator-level $\PQc$ quark with $\pt^{\PQc} > 25\GeV$ and $\abs{\eta^{\PQc}} < 2.5$,
and separated from the lepton by a distance $\Delta R(\PQc,\ell)>0.5$.
A prediction for the $\Rcpm$ ratio is similarly derived.
The \MADGRAPH prediction for the cross section is $\SWc = 110.9 \pm 0.2 \stat \unit{pb}$, and, for the cross section ratio, it is $\Rcpm$ = $0.969 \pm 0.004 \stat$. 
They are in agreement with the measured values within uncertainties. 

\section{Differential \texorpdfstring{$\Wc$}{W + c} cross section and \texorpdfstring{$(\PWpc)/(\PWmc)$}{(W+ + cbar)/(W- + c)} cross section ratio~\label{sec:xsec_diff}}

The $\Wc$ production cross section and $\Rcpm$ are measured differentially, as functions of $\abs{\etaell}$ and $\ptell$.
The binning of the differential distributions is chosen such that each bin is sufficiently populated to perform the measurement. 
Event migration between neighbouring bins caused by detector resolution effects is evaluated with the simulated signal sample and is negligible. 
The total sample is divided into subsamples according to the value of $\abs{\etaell}$ or $\ptell$, and the cross section and 
cross section ratio are computed using Eq.~(\ref{eq:W_c_data}). 
There is no significant dependence of the fraction of remaining background events, $f_{\text{bkg}}$, after \OSSS on $\abs{\etaell}$, 
whereas it decreases by a factor of two along the studied $\pt$ range.

The charm identification efficiency and its description in simulation vary with the $\pt$ of the jet containing the $\PQc$ quark. 
In $\Wc$ events, there is a correlation between the transverse momentum of the $\PQc$ jet and that of the lepton from the {\PW} boson decay. 
Thus, for the determination of the differential cross sections as a function of $\ptell$,
we apply charm identification efficiency scale factors, dependent on jet $\pt$, to the simulated samples.
These jet $\pt$-dependent scale factors are determined using the same procedure described in section~\ref{sec:Wsel_SV} 
by dividing the SL sample into subsamples depending on the jet $\pt$ and computing data-to-simulation scale factors for the efficiency of charm identification 
through the reconstruction of a secondary vertex for each of them. The value of the scale factors range from 0.9 to 1.0.

Systematic uncertainties in the differential $\Wc$ cross sections are in the range of 7--8\% for the SL channels and 4--5\% for the SV channels.
The main sources of the systematic uncertainty are related to the charm hadron decay rates in simulation, the charm identification efficiencies, 
and the limited event count of the simulated samples.
The largest uncertainty for the differential cross section as a function of the lepton $\pt$ (4--5\%) arises from the uncertainty in the  
charm identification efficiency scale factors.
The systematic uncertainty for the differential cross section ratios is in the range of 2--3\% for both channels, essentially coming
from the limited event count of the simulated samples.

The $\Wc$ differential cross sections, obtained after the combination of the measurements in the four channels, 
as functions of $\abs{\etaell}$ and $\ptell$ are presented in Tables~\ref{tab:averaged_norm_diff_xsec} and~\ref{tab:averagedpt_norm_diff_xsec}.
The combination of the differential $\Rcpm$ values is given in Table~\ref{tab:averaged_chargeratio_eta} 
as a function of $\abs{\etaell}$, and in Table~\ref{tab:averaged_chargeratio_pt} as a function of $\ptell$. 
The \convino tool is used for the combination; systematic uncertainties are assumed to be fully correlated among bins of the differential distributions.
\begin{linenomath*}
\begin{table}[htbp]
  \centering
    \topcaption{Measured differential cross section as a function of $\abs{\etaell}$, $\SWcdifflineeta$ from the combination of all four channels.}
    \renewcommand{\arraystretch}{1.2}
    \begin{tabular}{ccc}
     $[{\abs{\etaell}}_\text{min},{\abs{\etaell}}_\text{max}]$ & $\SWcdifflineeta$ [pb] \\
     \hline
     $[0.0,0.2]$ & $68.2 \pm 0.9\stat \pm 3.1\syst$ \\
     $[0.2,0.4]$ & $67.8 \pm 1.0\stat \pm 3.0\syst$ \\
     $[0.4,0.6]$ & $65.9 \pm 0.9\stat \pm 3.0\syst$ \\
     $[0.6,0.8]$ & $64.8 \pm 0.9\stat \pm 2.9\syst$ \\
     $[0.8,1.1]$ & $61.2 \pm 0.8\stat \pm 2.8\syst$ \\
     $[1.1,1.4]$ & $53.0 \pm 0.8\stat \pm 2.4\syst$ \\
     $[1.4,1.7]$ & $45.4 \pm 0.9\stat \pm 2.1\syst$ \\
     $[1.7,2.1]$ & $37.9 \pm 0.8\stat \pm 1.8\syst$ \\
    \end{tabular}
    \label{tab:averaged_norm_diff_xsec}
\end{table}
\end{linenomath*}

\begin{linenomath*}
\begin{table}[htbp]
  \centering
    \topcaption{Measured differential cross section as a function of $\ptell$, $\SWcdifflinept$ from the combination of all four channels.}
    \cmsTableii{
    \renewcommand{\arraystretch}{1.2}
    \begin{tabular}{cc}
     $[{\ptell}_\text{min},{\ptell}_\text{max}]$ [\GeVns{}] & $\SWcdifflinept$ [pb/\GeVns{}] \\
     \hline
     $[30,35]$   & $2.89  \pm 0.06 \stat \pm 0.15 \syst$  \\
     $[35,40]$   & $3.14  \pm 0.05 \stat \pm 0.16 \syst$  \\
     $[40,50]$   & $2.99  \pm 0.03 \stat \pm 0.15 \syst$  \\
     $[50,60]$   & $2.36  \pm 0.03 \stat \pm 0.12 \syst$  \\
     $[60,80]$   & $1.108  \pm 0.012 \stat \pm 0.055 \syst$  \\
     $[80,100]$  & $0.365 \pm 0.007\stat \pm 0.020 \syst$ \\
     $[100,200]$ & $0.0462 \pm 0.0014\stat \pm 0.0029 \syst$ \\
    \end{tabular}}
    \label{tab:averagedpt_norm_diff_xsec}
\end{table}
\end{linenomath*}

\begin{linenomath*}
 \begin{table}[htbp]
  \centering
   \topcaption{Measured cross section ratio $\Rcpm$ as a function of $\abs{\etaell}$, from the combination of all four channels.}
   \renewcommand{\arraystretch}{1.2}
   \begin{tabular}{c c}
    $[{\abs{\etaell}}_\text{min},{\abs{\etaell}}_\text{max}]$ & $\Rcpm$\\ 
    \hline
    $[0.0, 0.2]$ & $ 0.961 \pm 0.027\stat \pm 0.018\syst$ \\
    $[0.2, 0.4]$ & $ 1.003 \pm 0.030\stat \pm 0.021\syst$ \\
    $[0.4, 0.6]$ & $ 1.024 \pm 0.030\stat \pm 0.018\syst$ \\
    $[0.6, 0.8]$ & $ 0.982 \pm 0.029\stat \pm 0.023\syst$ \\
    $[0.8, 1.1]$ & $ 1.012 \pm 0.026\stat \pm 0.019\syst$ \\
    $[1.1, 1.4]$ & $ 1.019 \pm 0.030\stat \pm 0.020\syst$ \\
    $[1.4, 1.7]$ & $ 0.958 \pm 0.040\stat \pm 0.026\syst$ \\
    $[1.7, 2.1]$ & $ 0.874 \pm 0.037\stat \pm 0.027\syst$ \\
   \end{tabular}
   \label{tab:averaged_chargeratio_eta}
 \end{table}
\end{linenomath*}

\begin{linenomath*}
 \begin{table}[htbp]
  \centering
   \topcaption{Measured cross section ratio $\Rcpm$ as a function of $\ptell$, from the combination of all four channels.}
   \renewcommand{\arraystretch}{1.2}
   \begin{tabular}{c c}
    $[{\ptell}_\text{min},{\ptell}_\text{max}]$ [\GeVns{}] &  $\Rcpm$\\
    \hline
    $[30, 35]$  & $  0.893\pm 0.035\stat \pm 0.025\syst$ \\
    $[35, 40]$  & $  1.094\pm 0.039\stat \pm 0.034\syst$ \\
    $[40, 50]$  & $  1.006\pm 0.022\stat \pm 0.026\syst$ \\
    $[50, 60]$  & $  0.968\pm 0.021\stat \pm 0.019\syst$ \\
    $[60, 80]$  & $  0.934\pm 0.020\stat \pm 0.018\syst$ \\
    $[80, 100]$ & $  0.875\pm 0.037\stat \pm 0.021\syst$ \\
    $[100, 200]$& $  0.908\pm 0.056\stat \pm 0.031\syst$ \\
   \end{tabular}
   \label{tab:averaged_chargeratio_pt}
 \end{table}
\end{linenomath*}

\section{Comparison with theoretical predictions~\label{sec:theory}}

{\tolerance 800
The measured total and differential cross sections and cross section ratios are compared in this section with the analytical calculations 
from the $\MCFM$ 8.2 program~\cite{Campbell:1999ah,Campbell:2010ff}.
The $\Wc$ process description is available in $\MCFM$ up to $\mathcal{O}(\alpS^2)$ with a massive charm quark ($m_{\PQc}=1.5\GeV$).
The $\MCFM$ predictions for this process do not include contributions from gluon splitting into a $\ccbar$ pair, but only contributions
where the strange (or the down) quark couples to the {\PW} boson.
The implementation of the $\Wc$ process follows the calculation for the similar single top quark $\cPqt\PW$ process~\cite{MCFM_WplusC}. 
The parameters of the calculation are adjusted to match the experimental measurement: $\ptell>30\GeV$, $\abs{\etaell}<2.1$, 
$\pt^{\PQc}>25\GeV$, and $\abs{\eta^{\PQc}}<2.5$.
\par}

{\tolerance 800
We compute predictions for the following NLO PDF sets: MMHT2014~\cite{MMHT2014nlo}, CT14~\cite{CT14nlo}, NNPDF3.1 \cite{NNPDF31nlo}, and ABMP16~\cite{ABMP16nlo}.
They include dimuon data from neutrino-nucleus deep inelastic scattering to provide information on the strange quark content of the proton.
Both the factorization and the renormalization scales are set to the {\PW} boson mass, $m_{\PW}$.
To estimate the uncertainty from missing higher perturbative orders, cross section predictions are computed by varying
independently the factorization and renormalization scales to twice and half their nominal values, with the constraint that the ratio of the two scales is never larger than 2.
The envelope of the cross sections with these scale variations defines the theoretical scale uncertainty.
\par}

The value in the calculation of the strong coupling at the energy scale of the mass of the {\PZ} boson, $\alpS(m_{\PZ})$, 
is set to $\alpS(m_{\PZ}) = 0.118 (0.119)$ for the predictions with MMHT2014, CT14 and NNPDF3.1 (ABMP16).
Uncertainties in the predicted cross sections associated with $\alpS(m_{\PZ})$ are evaluated as half the difference in the predicted cross sections
evaluated with a variation of $\Delta(\alpS)=\pm 0.002$.
Uncertainties associated with the value of $\alpS(m_{\PZ})$ for the ABMP16 PDF set are given together with their PDF uncertainties and are not quoted separately in the tables.

\begin{figure*}[!tb]
 \centering
  \includegraphics[width=0.48\textwidth]{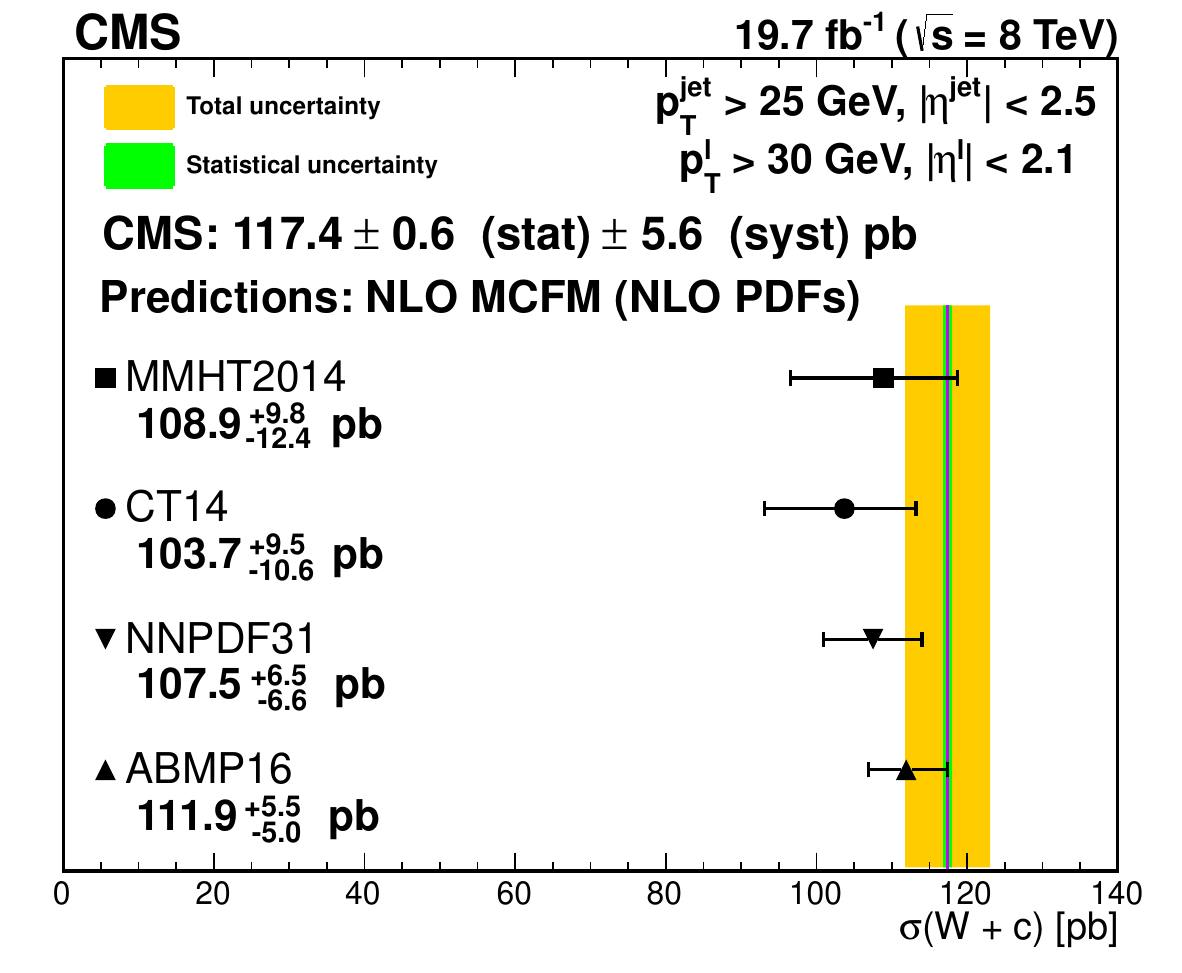}
  \includegraphics[width=0.48\textwidth]{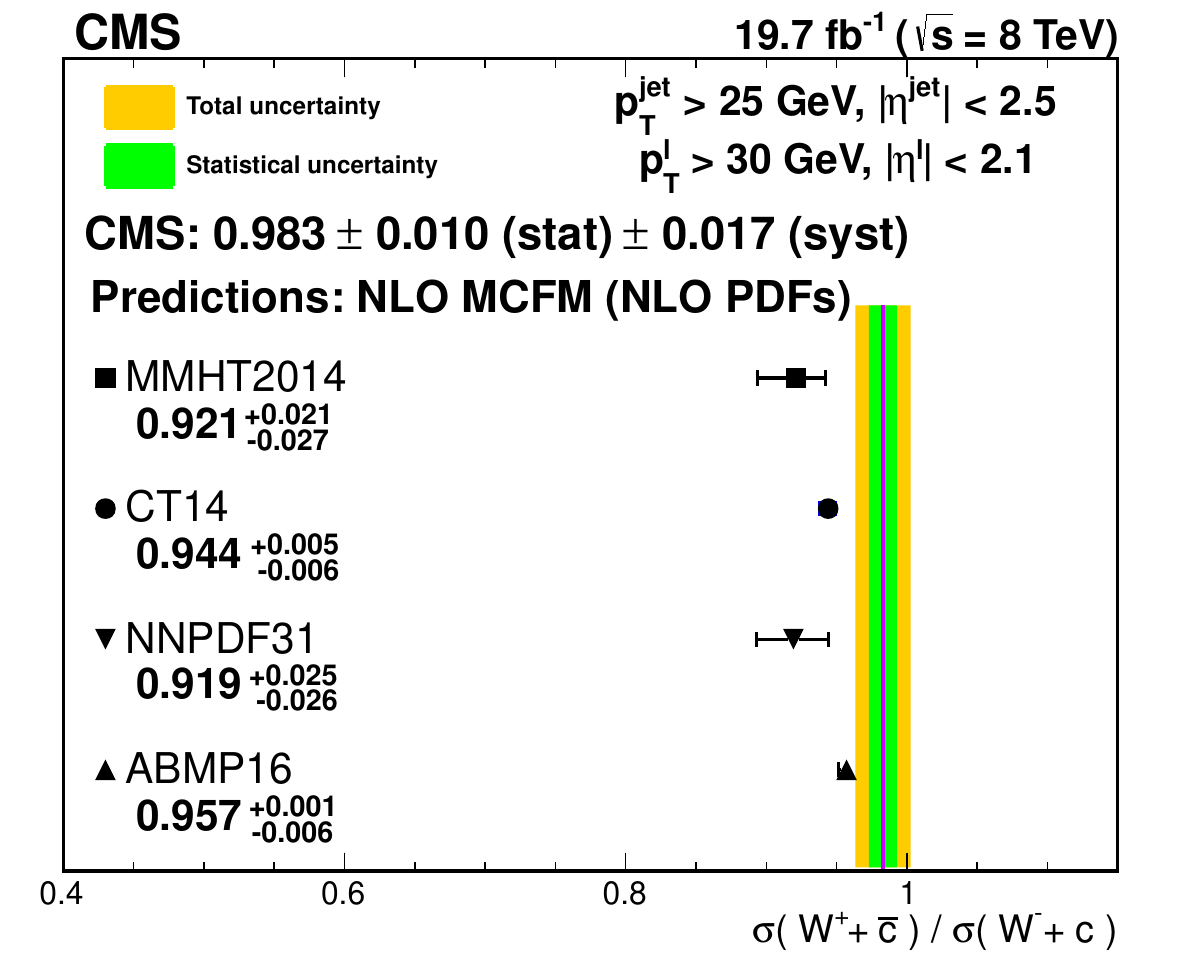}
  \caption{Comparison of the theoretical predictions for $\SWc$ (left) and $\SWpc/\SWmc$ (right) computed with $\MCFM$ and several sets of PDFs 
  with the current experimental measurements.}
 \label{fig:summary}
\end{figure*}

The theoretical predictions for the fiducial $\Wc$ cross section 
are summarized in Table~\ref{table:MCFM_Sc}, where the central value of each prediction
is given, together with the uncertainty arising from the PDF variations within each set, the choice of scales, and $\alpS$.
The experimental result reported in this paper is also included in Table~\ref{table:MCFM_Sc}.
The size of the PDF uncertainties depends on the different input data and methodology
used by the various groups. In particular, they depend on the parameterization of the strange quark PDF and on the definition of the one standard deviation
uncertainty band. The maximum difference between the central values of the various PDF predictions is ${\sim}8\%$. 
This difference is smaller than the total uncertainty in each of the individual predictions. 
Theoretical predictions are in agreement within the uncertainties with the measured cross section, as depicted in Fig.~\ref{fig:summary} (left), 
although lower. 

Theoretical predictions for $\SWpc$ and $\SWmc$ are computed independently 
in the same phase space of the measurement under the same conditions previously explained.
Expectations for $\Rcpm$ are derived from them and presented in Table~\ref{table:MCFM_Rcpm}.
All theoretical uncertainties are significantly reduced in the cross section ratio prediction.
The theoretical predictions of the cross section ratio agree with each other, with the largest difference reaching 4\%.
The experimental value is larger than the theoretical predictions, but it is within two or three standard deviations depending
on the prediction.
They are presented graphically in Fig.~\ref{fig:summary} (right).
The ratio of cross sections is sensitive to the asymmetry in the strange quark-antiquark content in the proton,
but also to the down quark and antiquark asymmetry from the Cabibbo-suppressed process
$\dbar\Pg \to \PWpc$ ($\PQd\Pg \to \PWmc$). 
The $\PQd$-$\dbar$ asymmetry is larger in absolute value than
the difference between strange quarks and antiquarks.
It is worth noting that the CT14 PDF theoretical predictions assumes no strangeness asymmetry.
\begin{table*}[htbp]
 \centering
  \topcaption{Theoretical predictions for $\SWc$ from $\MCFM$ at NLO.
  The kinematic selection follows the fiducial phase space definition: $\ptell>30\GeV$, $\abs{\etaell}<2.1$, $\pt^{\PQc}>25\GeV$, $\abs{\eta^{\PQc}}<2.5$, and $\Delta R(\PQc,\ell)>0.5$.
  For each PDF set, the central value of the prediction is given, together with the
  relative uncertainty as prescribed from the PDF set, and the uncertainties associated with the scale variations and with the value of \alpS.
  The total uncertainty is given in the last column.
  The last row in the table gives the experimental results presented in this paper.}
  \renewcommand{\arraystretch}{1.2}
  \begin{tabular}{cccccc}
   PDF set   & $\SWc$ [pb] & $\delta_{\text{PDF}} [\%]$ & $\delta_{\text{scales}} [\%]$ & $\delta_{\alpS} [\%]$ & Total uncert. [pb] \\
   \hline
   MMHT2014  & $108.9$ & $^{+6.0}_{-9.1}$ & $^{+4.4}_{-4.6}$ & $\pm5$ & $^{\phantom{1}+9.8}_{-12.4}$  \\
   CT14      & $103.7$ & $^{+7.6}_{-8.7}$ & $^{+4.5}_{-4.6}$ & $\pm2.2$ & $^{\phantom{1}+9.5}_{-10.6}$ \\
   NNPDF3.1  & $107.5$ & $\pm 3.5$ & $^{+4.4}_{-4.5}$ & $\pm2.2$ & $^{\phantom{1}+6.5}_{\phantom{1}-6.6}$  \\ 
   ABMP16    & $111.9$ & $\pm 0.9$ & $^{+4.8}_{-4.4}$ & \NA  & $^{\phantom{1}+5.5}_{\phantom{1}-5.0}$ \\ 
   \hline
   CMS       & \multicolumn{5}{c}{$117.4 \pm 0.6 {\stat} \pm 5.6 {\syst}$ \unit{pb}}  \\
  \end{tabular}
  \label{table:MCFM_Sc}
\end{table*}
\begin{table*}[htbp]
 \centering
  \topcaption{Theoretical predictions for $\Rcpm$ calculated with $\MCFM$ at NLO.
  The kinematic selection follows the experimental requirements: $\ptell>30\GeV$, $\abs{\etaell}<2.1$, $\pt^{\PQc}>25\GeV$, $\abs{\eta^{\PQc}}<2.5$, and $\Delta R(\PQc,\ell)>0.5$.
  For each PDF set, the central value of the prediction is given, together with the
  relative uncertainty as prescribed from the PDF set, and the uncertainties associated with the scale variations and with the value of $\alpS$.
  The total uncertainty is given in the last column.
  The last row in the table gives the experimental results presented in this paper.}
  \renewcommand{\arraystretch}{1.2}
  \begin{tabular}{cccccc}
   PDF set   & $\Rcpm$ & $\delta_{\text{PDF}} [\%]$ & $\delta_{\text{scales}} [\%]$ & $\delta_{\alpS} [\%]$ & Total uncert.\\
   \hline
   MMHT2014  & $0.921$ & $^{+2.2}_{-2.8}$ & $^{+0.3}_{-0.2}$ & $\pm0.3$  & $^{+0.021}_{-0.027}$\\
   CT14      & $0.944$ & $^{+0.4}_{-0.6}$ & $^{+0.4}_{-0.2}$ & $\pm0.1$ & $^{+0.005}_{-0.006}$ \\
   NNPDF3.1  & $0.919$ & $\pm 2.6$ & $^{+0.1}_{-0.6}$ & $\pm0.8$ & $^{+0.025}_{-0.026}$ \\ 
   ABMP16    & $0.957$ & $\pm 0.1$ & $^{+0.0}_{-0.7}$ & \NA & $^{+0.001}_{-0.006}$ \\ 
   \hline
   CMS       & \multicolumn{5}{c}{$0.983 \pm 0.010 {\stat} \pm 0.017 {\syst}$}  \\
  \end{tabular}
  \label{table:MCFM_Rcpm}
\end{table*}

Predictions for the differential cross sections are obtained from analytical calculations with $\MCFM$,
using the same binning as in the data analysis. Systematic uncertainties in the scale variations in some pseudorapidity bins and for some PDF sets reach 10\%.
Scale uncertainties in the differential cross sections as a function of $\ptell$ are larger than in those as a function of $\abs{\etaell}$.

\begin{figure}[htbp]
 \centering
     \includegraphics[width=0.48\textwidth]{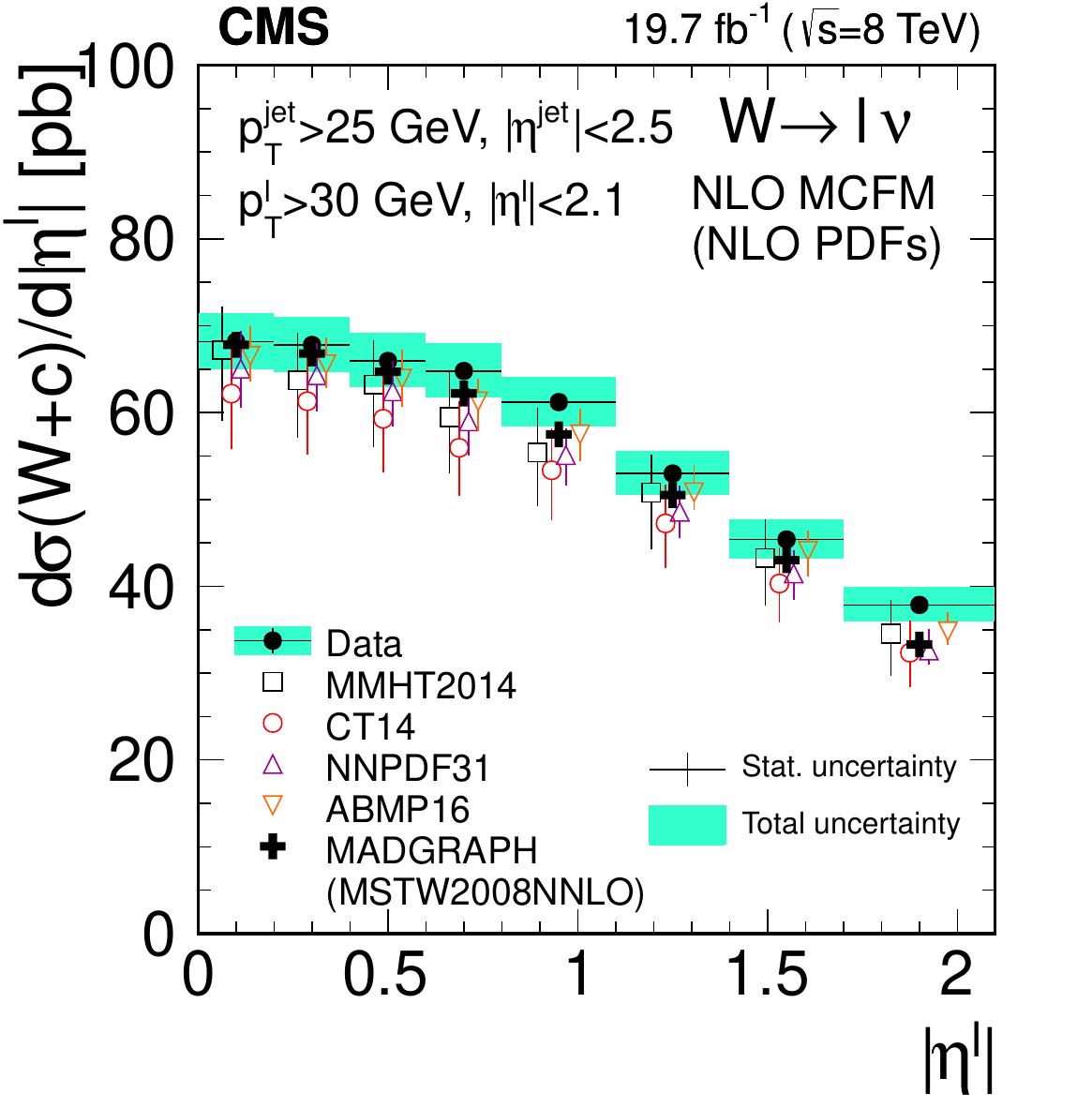}
     \includegraphics[width=0.48\textwidth]{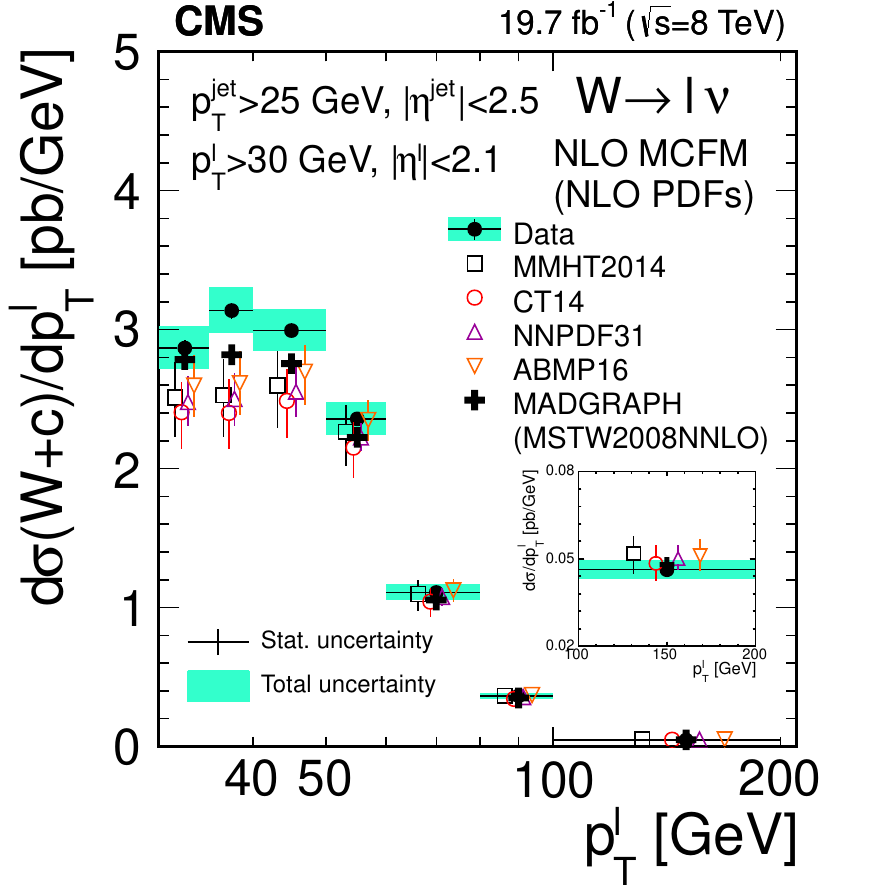}
     \caption{Differential cross sections, $\SWcdifflineeta$ (\cmsLeft) and $\SWcdifflinept$ (\cmsRight).
              The data points are the combination of the results with the four different samples: SL  and SV samples in $\Wen$ and $\Wmn$ events.
              Theoretical predictions at NLO computed with $\MCFM$ and four different NLO PDF sets are also shown.
              Symbols showing the theoretical expectations are slightly displaced in the horizontal axis for better visibility.
              The error bars in the $\MCFM$ predictions include PDF, \alpS, and scale uncertainties.
              The inset in the \cmsRight plot, $\SWcdifflinept$, zooms into the measurement-prediction comparison for the last bin, $100<\ptell<200\GeV$.
              Predictions from \MADGRAPH using the PDF set MSTW2008NNLO are also presented.}
  \label{fig:Sc_w_th}
\end{figure}

{\tolerance 820
The theoretical predictions are compared with the combination of the experimental measurements presented in Section~\ref{sec:xsec_diff}.
Figure~\ref{fig:Sc_w_th} shows the measurements given in 
Tables~\ref{tab:averaged_norm_diff_xsec} and~\ref{tab:averagedpt_norm_diff_xsec}, and predictions for the differential cross sections as functions
of $\abs{\etaell}$ and $\ptell$, respectively. 
Theoretical predictions from \MADGRAPH using the PDF set MSTW2008NNLO are also shown. 
The shape of the differential distribution as a function of $\abs{\etaell}$ is well described by all theoretical predictions.
Theoretical predictions are about 10\% lower than the measured cross section in the low transverse momentum region, $\ptell<50\GeV$. Recent calculations~\cite{czakon2020nnlo} point to NNLO corrections between 5 and 10\% that bring theoretical predictions closer to the measurements.   
\par}

\begin{figure}[htbp]
 \centering
     \includegraphics[width=0.48\textwidth]{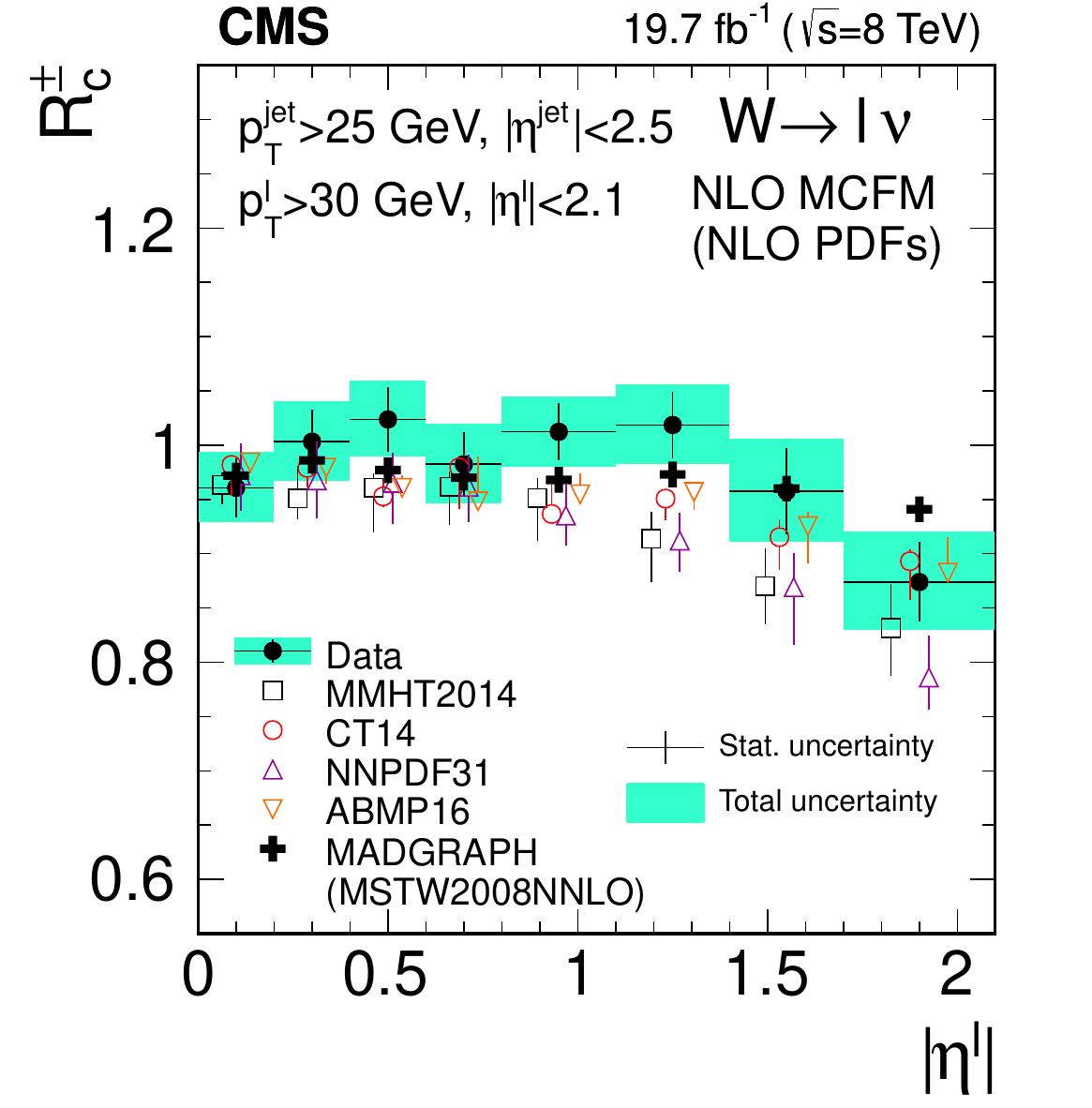}
     \includegraphics[width=0.48\textwidth]{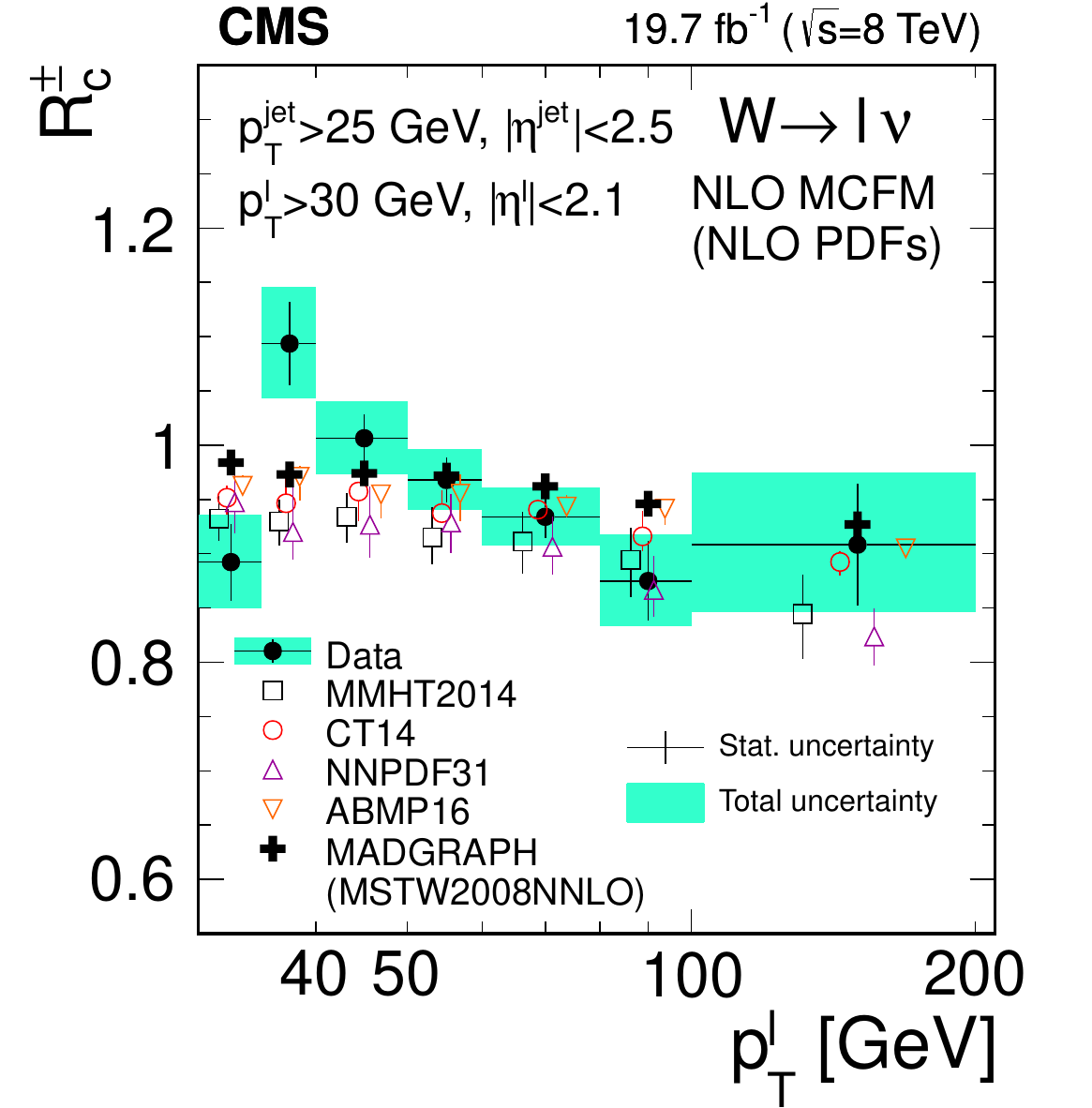}
     \caption{Cross section ratio, $\Rcpm$, as functions of $\abs{\etaell}$ (\cmsLeft) and $\ptell$ (\cmsRight).
              The data points are the combination of the results from the SL and SV samples in $\Wen$ and $\Wmn$ events.
              Theoretical predictions at NLO computed with $\MCFM$ and four different NLO PDF sets are also shown.
              Symbols showing the theoretical expectations are slightly displaced in the horizontal axis
              for better visibility.
              The error bars in the $\MCFM$ predictions include PDF, \alpS, and scale uncertainties.
              Predictions from \MADGRAPH using the PDF set MSTW2008NNLO are also presented.}
     \label{fig:Rpm_w_th}
\end{figure}

The predictions for the differential cross section ratio as functions of $\abs{\etaell}$ and $\ptell$ are presented in Fig.~\ref{fig:Rpm_w_th}, 
together with the cross section ratios given in Tables~\ref{tab:averaged_chargeratio_eta} and~\ref{tab:averaged_chargeratio_pt}.
Theoretical predictions from \MADGRAPH are also shown.
The measured cross section ratio, as a function of $\ptell$, is larger than the predictions in the 35--60\GeV range but compatible within uncertainties. 
According to Ref.~\cite{czakon2020nnlo}, NNLO corrections for $\ptell<60\GeV$ are of the order of 5\%, and are around 1\% for $\ptell>60\GeV$.
These corrections would improve the description of the measurements in the low $\ptell$ region.    

\section{Impact on the strange quark distribution determination}\label{sec:qcd_analysis}

The associated $\Wc$ production at a centre-of-mass energy of 8\TeV directly probes the strange quark distribution of the proton at the scale of 
$m^2_{\PW}$, in the kinematic range of $0.001< x <0.080$, 
where $x$ is the fraction of the proton momentum taken by the struck parton in the infinite-momentum frame. 
The present combined measurement of the $\Wc$ production cross section, determined as a function of 
$\abs{\etaell}$ and for lepton $\ptell>30\GeV$, is used in a QCD analysis at NLO. 

The combination of the HERA inclusive deep inelastic scattering (DIS) cross sections~\cite{Abramowicz:2015mha} and the 
available CMS measurements of the lepton charge asymmetry in {\PW} boson production at $\rts=7$ and 8\TeV~\cite{Chatrchyan:2013mza, Khachatryan:2016pev} are used. 
The CMS measurements probe the valence quark distributions in the kinematic range $10^{-3} \leq x \leq 10^{-1}$ and have indirect sensitivity to the strange quark distribution. 
The CMS measurements of $\Wc$ production at $\rts=7$~\cite{CMS-PAPER-SMP-12-002} and 13\TeV~\cite{CMS-PAPER-SMP-17-014} 
are also used in a joint QCD analysis to fully exploit the other measurements at CMS that are sensitive to the strange quark distribution.
The measurements included in this analysis are the HERA combined reduced cross sections for charged and neutral currents as 
a function of $Q^2$ and $x$ for different centre-of-mass energies, the muon charge asymmetry as a function of the pseudorapidity of the muon,
 and the $\Wc$ differential cross section as a function of $\abs{\etaell}$.

The correlations of the experimental uncertainties for each individual data set are included.
The systematic uncertainties in the semileptonic branching fraction are treated as correlated
between the CMS measurements of $\Wc$ production at 7 and 8\TeV.
The rest of the systematic uncertainties are treated as uncorrelated between the two data-taking periods.
The measurements of $\Wc$ production at a centre-of-mass energy of 13\TeV are treated as uncorrelated with those at 7 and 8\TeV 
because of the different methods of charm tagging and the differences in reconstruction and event selection in these data sets.

{\tolerance 800
The theoretical predictions for the muon charge asymmetry and for the $\Wc$ production are calculated at NLO using 
the \MCFM 6.8 program~\cite{Campbell:1999ah,Campbell:2010ff}, which is interfaced with \mbox{\textsc{applgrid}} 1.4.56~\cite{Carli:2010rw}. 
The open-source QCD fit framework for PDF determination \mbox{\textsc{xFitter}}~\cite{Alekhin:2014irh, herafitter}, version 2.0.0, 
is used with the parton distributions evolved using the Dokshitzer--Gribov--Lipatov--Altarelli--Parisi 
equations~\cite{Gribov:1972ri,Altarelli:1977zs,Curci:1980uw,Furmanski:1980cm,Moch:2004pa,Vogt:2004mw} at NLO, 
as implemented in the \mbox{\textsc{qcdnum 17-00/06}} program~\cite{Botje:2010ay}. 
The Thorne--Roberts \cite{Thorne:2006qt,Martin:2009iq} general mass variable flavour number scheme at NLO is used for 
the treatment of heavy quark contributions with heavy quark masses $m_{\PQb} = 4.5\GeV$ and $m_{\PQc} = 1.5\GeV$, 
which correspond to the values used in the signal MC simulation in the cross section measurements.
The renormalization and factorization ($\mu_f$) scales are set to $Q$, which denotes the four-momentum transfer in the case of 
the DIS data and $m_{\PW}$ in the case of the muon charge asymmetry and the $\Wc$ process.
The strong coupling is set to $\alpS(m_{\PZ})$ = 0.118. The $Q^2$ range of the HERA data is restricted to 
$Q^2 \geq Q^2_{\min} = 3.5\GeV^2$ to ensure the applicability of perturbative QCD over the kinematic range of the fit. 
The procedure for the determination of the PDFs follows that of Ref.~\cite{CMS-PAPER-SMP-17-014}.
\par}

The PDFs of the proton, $xf(x)$, are generically parameterized at the starting scale 
\begin{linenomath*}
 \begin{equation}
   xf(x) = A x^{B} (1-x)^{C} (1 + D x + E x^2).
   \label{eqn:pdf}
 \end{equation}
\end{linenomath*}
The parameterized PDFs are the gluon distribution, $x\Pg$, the valence quark distributions, $x\PQu_\mathrm{v}$, $x\PQd_\mathrm{v}$, 
the $\PQu$-type and $\PQd$-type anti-quark distributions, $x\ubar$, $x\dbar$, and $x\PQs$ ($x\sbar$) denoting the strange (anti-)quark distribution.
By default it is assumed that $x\PQs=x\sbar$.

The central parameterization at the initial scale of the QCD evolution chosen as $Q^2_{0} = 1.9\GeV^2$ is
\begin{linenomath*}
\begin{eqnarray}
\label{eq:xgpar}
x\Pg(x) &=   & A_{\Pg} x^{B_{\Pg}} (1-x)^{C_{\Pg}}  ,  \\
\label{eq:xuvpar}
x\PQu_\mathrm{v}(x) &=  & A_{\PQu_\mathrm{v}} x^{B_{\PQu_\mathrm{v}}}  (1-x)^{C_{\PQu_\mathrm{v}}}\left(1+E_{\PQu_\mathrm{v}}x^2 \right) , \\
\label{eq:xdvpar}
x\PQd_\mathrm{v}(x) &=  & A_{\PQd_\mathrm{v}} x^{B_{\PQd_\mathrm{v}}}  (1-x)^{C_{\PQd_\mathrm{v}}} , \\
\label{eq:xubarpar}
x\ubar(x) &=  & A_{\ubar} x^{B_{\ubar}} (1-x)^{C_{\ubar}}\left(1+D_{\ubar}x\right) , \\
\label{eq:xdbarpar}
x\dbar(x) &= & A_{\dbar} x^{B_{\dbar}} (1-x)^{C_{\dbar}} ,\\
\label{eq:xsbarpar}
x\sbar(x) &= & A_{\sbar} x^{B_{\sbar}} (1-x)^{C_{\sbar}}.
\end{eqnarray}
\end{linenomath*}

The parameters $A_{\PQu_\mathrm{v}}$ and $A_{\PQd_\mathrm{v}}$ are determined using 
the quark counting rules and $A_{\Pg}$ using the momentum sum rule~\cite{BasicQCD}.
The normalization and slope parameters, $A$ and $B$,
of $\ubar$ and $\dbar$  are set equal such that 
$x\ubar = x\dbar$ at very small $x$. 
The strange quark PDF $x\sbar$ is parameterized as in Eq.~(\ref{eq:xsbarpar}), 
with $B_{\sbar} = B_{\dbar}$, leaving 
two free strangeness parameters, $A_{\sbar}$ and $C_{\sbar}$.  
The optimal central parameterization was determined in a so-called parameterization scan 
following the HERAPDF procedure~\cite{Abramowicz:2015mha}.

For all measured data, the predicted and measured cross sections together with their corresponding uncertainties 
are used to build a global $\chi^2$, minimized to determine the initial PDF parameters~\cite{Alekhin:2014irh, herafitter}. 
The quality of the overall fit can be judged based on the global $\chi^2$ divided by the number of degrees of freedom, 
$n_{\mathrm{dof}}$. For each data set included in the fit, a partial $\chi^2$ divided by the number of measurements 
(data points), $n_{\mathrm{dp}}${\,}, is provided. The correlated part of $\chi^2$ reports on the influence of the correlated 
systematic uncertainties in the fit. 
The logarithmic penalty $\chi^2$ part comes from a $\chi^2$ term used to minimize bias.
The full form of the $\chi^2$  used in this analysis follows the HERAPDF2.0 analysis~\cite{Abramowicz:2015mha}. 
The global and partial $\chi^2$ values for each data set are listed in 
Table~\ref{chi2_paper_table_newparam}, illustrating a general agreement among all the data sets. 
The somewhat high $\chi^2$ values for the combined DIS data are very similar to those observed 
in Ref.~\cite{Abramowicz:2015mha}, where they are investigated in detail. 
The same fit, using the four different analysis channels instead of the combined measurement for $\Wc$ at $\rts=8\TeV$,
gives very consistent results and comparable values of $\chi^2$ for all data sets included.
\begin{linenomath*}
\begin{table*}[htbp]
  \centering
  \topcaption {The partial $\chi^2$ per number of data points, $n_{\mathrm{dp}}$, and the global $\chi^2$ per number of 
degrees of freedom, $n_{\mathrm{dof}}$, resulting from the PDF fit.}
  \renewcommand{\arraystretch}{1.2}
   \begin{tabular}{lllll} 
     Data set &   &  $\chi^2 / n_{\mathrm{dp}}$  \\ 
     \hline
     HERA I+II charged current & $\Pep \Pp$, $E_{\Pp}$ = 920\GeV & 41 / 39    \\ 
     HERA I+II charged current & $\Pem \Pp$, $E_{\Pp}$ = 920\GeV & 59/ 42     \\ 
     HERA I+II neutral current & $\Pem \Pp$, $E_{\Pp}$ = 920\GeV & 220 / 159  \\ 
     HERA I+II neutral current & $\Pep \Pp$, $E_{\Pp}$ = 820\GeV & 69 / 70    \\ 
     HERA I+II neutral current & $\Pep \Pp$, $E_{\Pp}$ = 920\GeV & 445 / 377  \\ 
     HERA I+II neutral current & $\Pep \Pp$, $E_{\Pp}$ = 460\GeV & 217 / 204  \\ 
     HERA I+II neutral current & $\Pep \Pp$, $E_{\Pp}$ = 575\GeV & 220 / 254  \\ 
     CMS {\PW} muon charge asymmetry 7\TeV (4.7\fbinv)& & 13.5 / 11  \\ 
     CMS {\PW} muon charge asymmetry 8\TeV (18.8\fbinv)& & 3.8 / 11   \\ 
     CMS {\Wc} 7\TeV (5\fbinv)& & 2.9 / 5  \\ 
     CMS {\Wc} 13\TeV (35.7\fbinv)& & 2.8 / 5&   \\ 
     CMS {\Wc} 8\TeV (19.7\fbinv)& & 3.0 / 8&   \\
     Correlated $\chi^2$ &  & 86  \\
     Log penalty $\chi^2$ &  & 5  \\  [1.ex] 
     Total $\chi^2/n_{\mathrm{dof}}$ &  & 1387 / 1171  
   \end{tabular}
  \label{chi2_paper_table_newparam}
\end{table*}
\end{linenomath*}

{\tolerance 800
The experimental PDF uncertainties are in\-ves\-ti\-gated according to the general approach of 
\textsc{HERAPDF}~\cite{Aaron:2009aa, Abramowicz:2015mha}. 
A cross check was performed using the MC method \cite{Giele:1998gw, Giele:2001mr}.
The parton distributions and their uncertainties obtained from both methods are consistent.
\par}

{\tolerance 800
We show results for the strange quark distribution $x\PQs(x,\PGm_f^2)$ and the strangeness suppression factor 
$R_{\PQs}(x,\PGm_f^2)$ = $(\PQs+\PAQs)/(\PAQu+\PAQd)$.
To investigate a possible impact of the assumptions on model input on the PDFs, alternative fits are performed, 
in which the heavy quark masses are set to $m_{\PQb} = 4.25$ and $4.75\GeV$, $m_{\PQc} = 1.45$ and $1.55\GeV$, 
and the value of $Q^2_\text{min}$ imposed on the HERA data is set to $2.5$ and $5.0\GeV^2$. 
These variations do not alter results on $x\PQs(x,\PGm_f^2)$ or $R_{\PQs}(x,\PGm_f^2)$ significantly, 
compared to the experimental PDF fit uncertainty.
\par}
The differences between the central fit and the fits corresponding  
to the variations of $Q^2_{\min}$, $m_{\PQc}$, and $m_{\PQb}$ are
added in quadrature, separately for positive and negative deviations, and 
represent the model uncertainty. 
The parameterization variations considered
 consist of adding extra $D$ and $E$ parameters in the polynomials of Eq.\,(\ref{eqn:pdf})
and varying the starting scale: $Q^2_{0}=1.6$ and $2.2\GeV^2$. 
In addition, further variations of the low-$x$ sea 
quark parameterization are allowed:
the $A$ and $B$ parameters for $\ubar$ and $\dbar$ are allowed to differ. 
The strange quark distribution and strangeness suppression factor are consistent with the nominal fit.
The parameterization uncertainty corresponds to the envelope of the fits described above.
The additional release of the condition $B_{\sbar}=B_{\dbar}$ in the fit results in a shape of the $\PQs$ quark PDF 
that could possibly violate the nonsinglet octet combination rules of QCD~\cite{Kataev:2003xp}. 
Therefore this fit is only used for the parameterization variation and not as a nominal fit.
The total PDF uncertainty is obtained by adding in quadrature 
the experimental, model, and parameterization uncertainties.

To assess the impact of the \Wc~ data collected at $\rts=8\TeV$ on $x\PQs(x,\PGm_f^2)$
and $R_{\PQs}(x,\PGm_f^2)$, another QCD fit 
is performed, using the same parameterization described in Eqs.~(\ref{eq:xgpar}--\ref{eq:xsbarpar})
but without these data. 
The central values of all parton distributions in those two fits are consistent within experimental 
uncertainties.
The results of these two QCD fits for the $\PQs$ quark PDF and 
$R_{\PQs}$ at the scale of $m^2_{\PW}$ are shown in Fig.~\ref{s_rs_impact}.
The relative total uncertainties are also compared in Fig.~\ref{s_rs_impact}.
The reduction of the uncertainties for these distribution with respect to those obtained without 
the new data is clearly visible.
\begin{linenomath*}
 \begin{figure*}[htb!]
  \centering
  \includegraphics[width=0.48\textwidth]{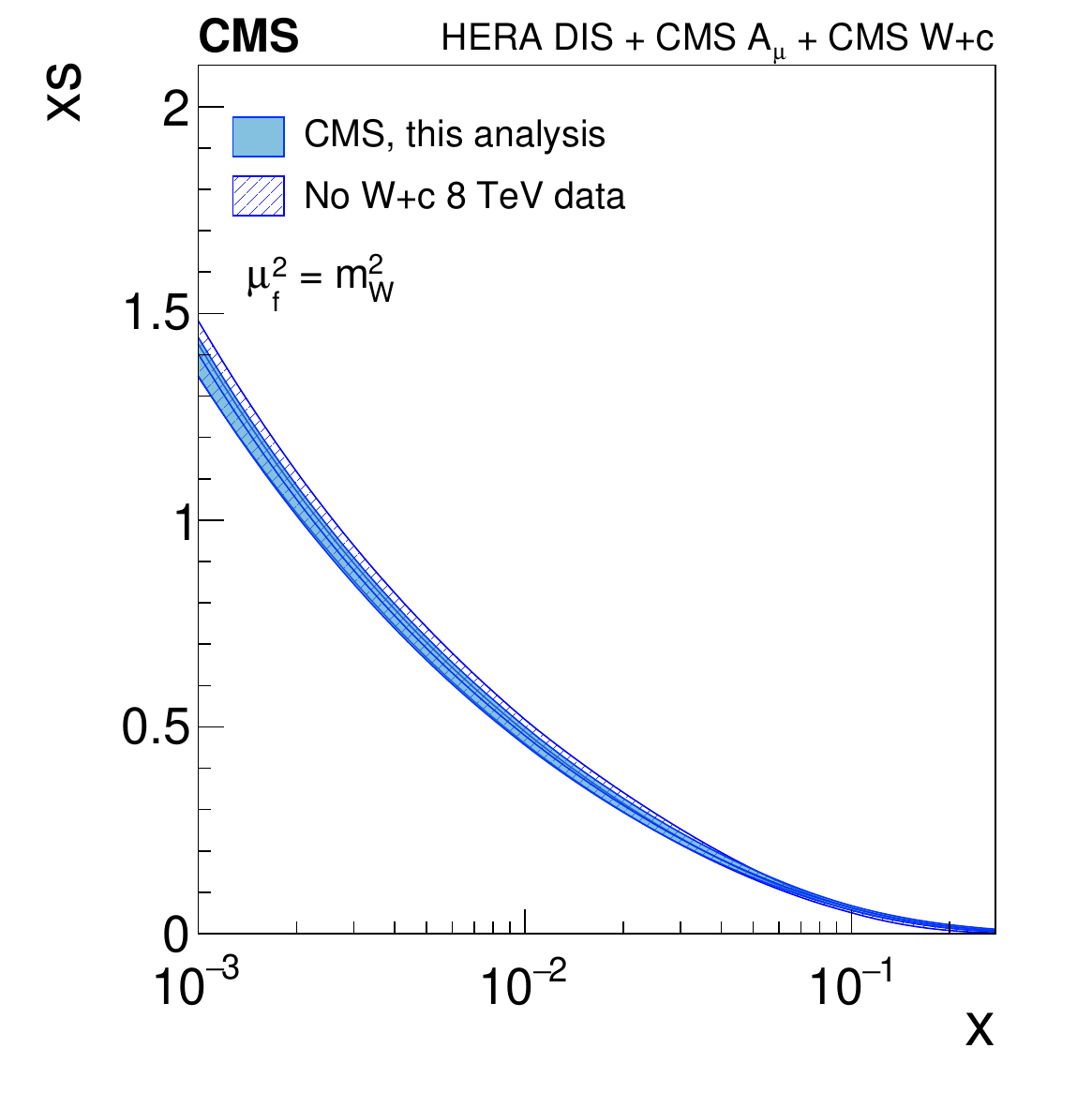}
  \includegraphics[width=0.48\textwidth]{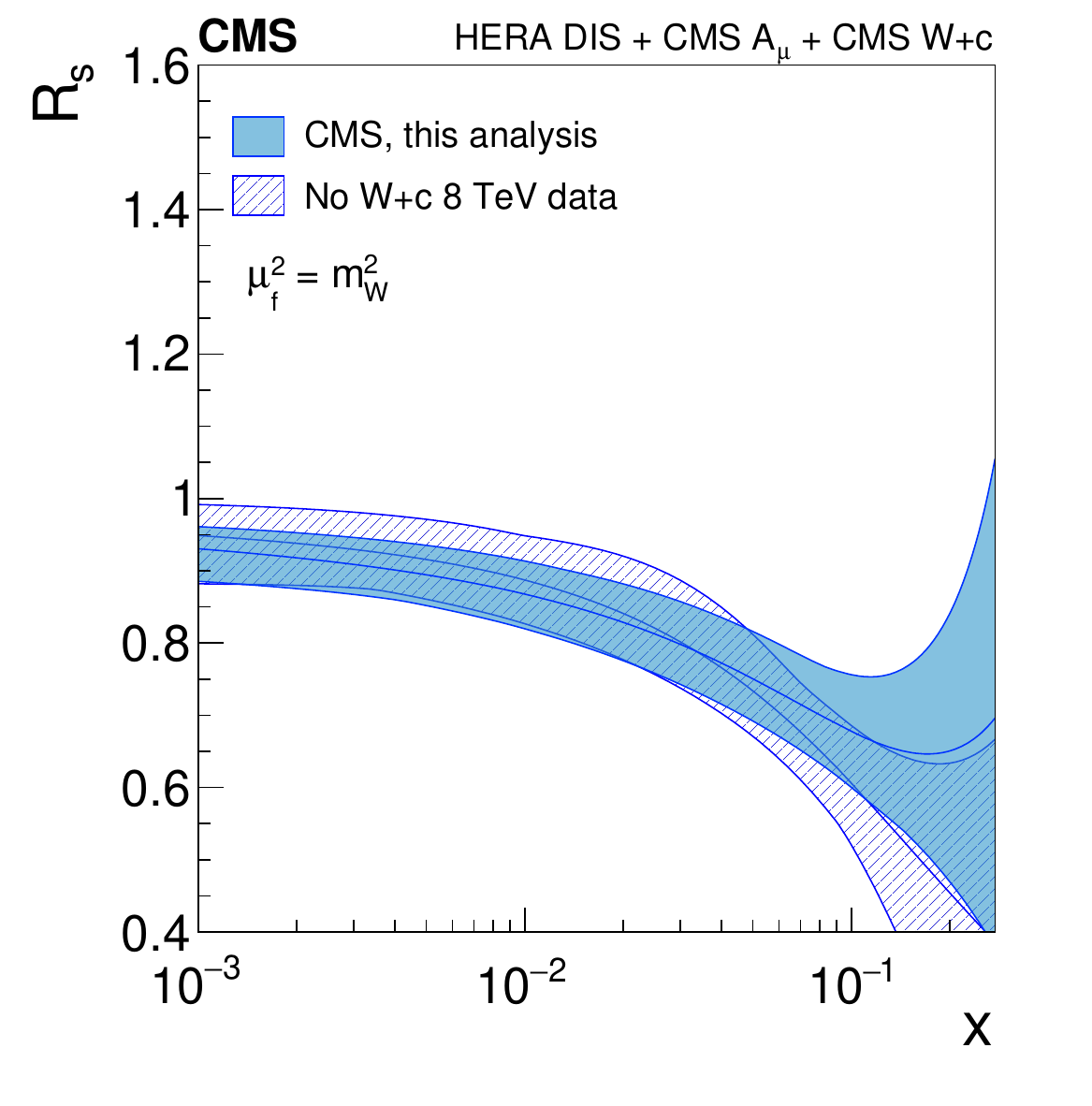}
  \includegraphics[width=0.48\textwidth]{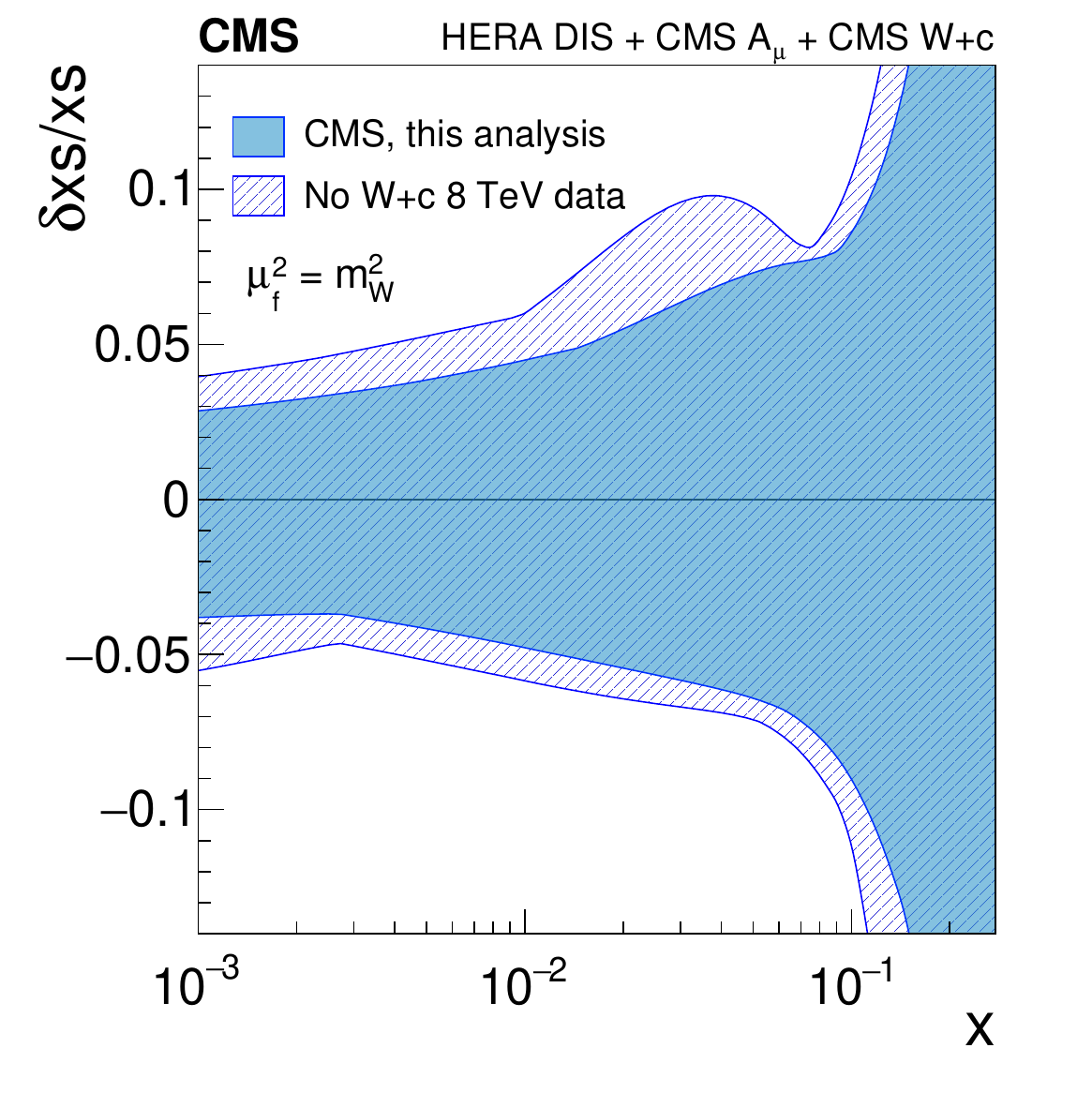}
  \includegraphics[width=0.48\textwidth]{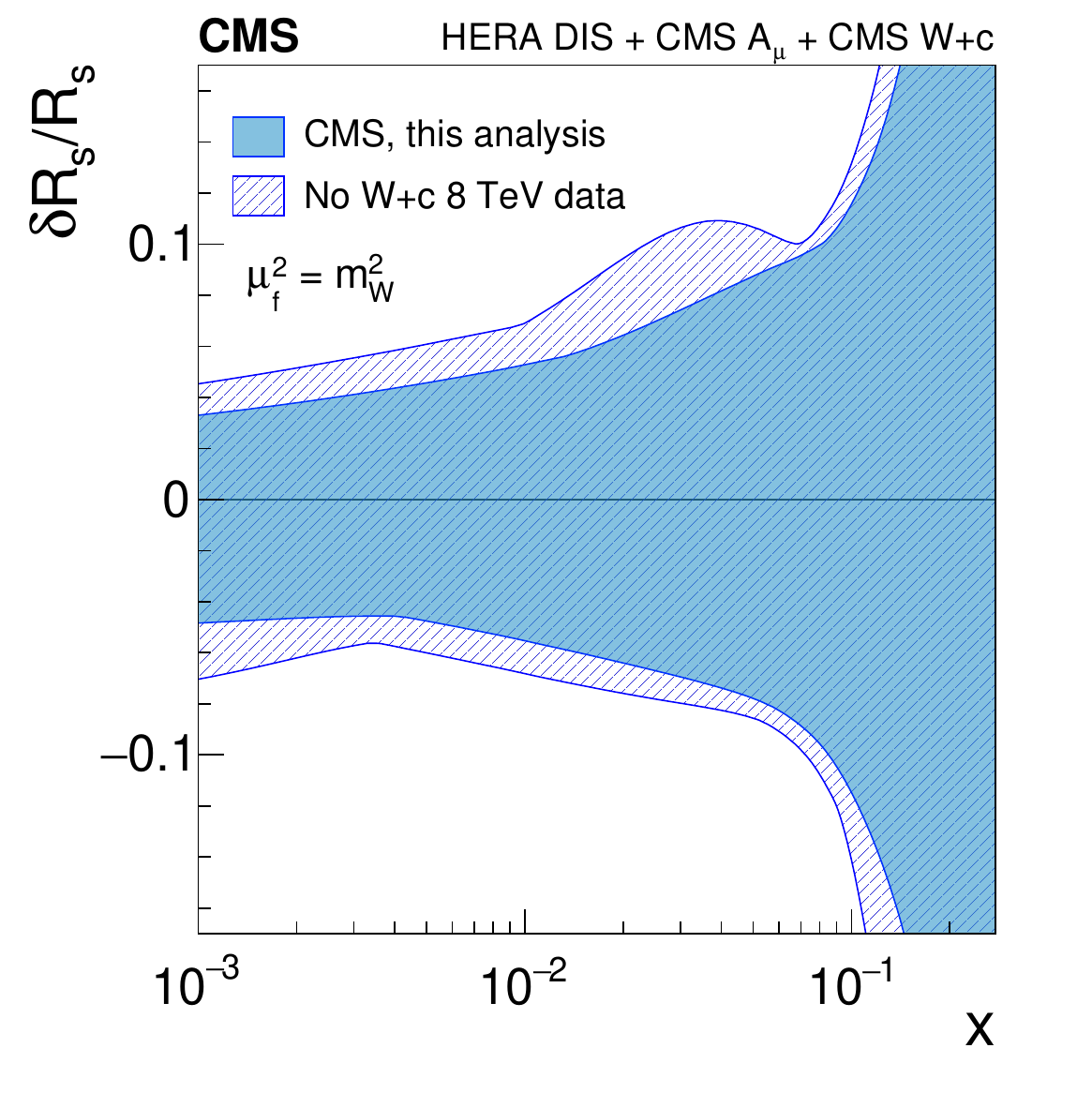}
  \caption{The strange quark distribution (upper left) and the strangeness suppression factor (upper right) as a function
  of $x$ at the factorization scale of $m^2_{\PW}$. The corresponding relative total uncertainties are compared in the lower plots 
 (strange quark distribution, lower left, and strangeness suppression factor, lower right).
  The results from the QCD analysis, shown as a filled area, use as input the combination of the inclusive deep inelastic scattering (DIS) 
  cross sections~\cite{Abramowicz:2015mha}, the CMS measurements of the lepton charge asymmetry in {\PW} boson production at 
  $\rts=7$ and 8\TeV~\cite{Chatrchyan:2013mza, Khachatryan:2016pev}, and the CMS measurements of $\Wc$ production at $\rts=7$~\cite{CMS-PAPER-SMP-12-002},
  ~8 (this analysis) and 13\TeV~\cite{CMS-PAPER-SMP-17-014}. The $\Wc$ measurement at $\rts=8\TeV$ is not used for the fit shown in hatched style.}
  \label{s_rs_impact}
 \end{figure*}
\end{linenomath*}

In Fig.~\ref{s_rs_pdfs}, the distributions of $x\PQs(x,\PGm_f^2)$ and
$R_{\PQs}(x,\PGm_f^2)$ at the scale of $m^2_{\PW}$ obtained in this analysis are presented together with 
the results of other global PDFs: 
ABMP16~\cite{ABMP16nlo},
NNPDF3.1~\cite{NNPDF31nlo},
CT18~\cite{CT18nlo},
and MSHT20 \cite{MSHT20nlo}.
These PDF sets have in common the use of the combined HERA data set, and also include 
neutrino charm production data and LHC {\PW} and {\PZ}  boson measurements to provide 
information on the strange quark content of the proton. 
The overall agreement between the various results is good.
\begin{linenomath*}
 \begin{figure*}[htb!]
  \centering
  \includegraphics[width=0.48\textwidth]{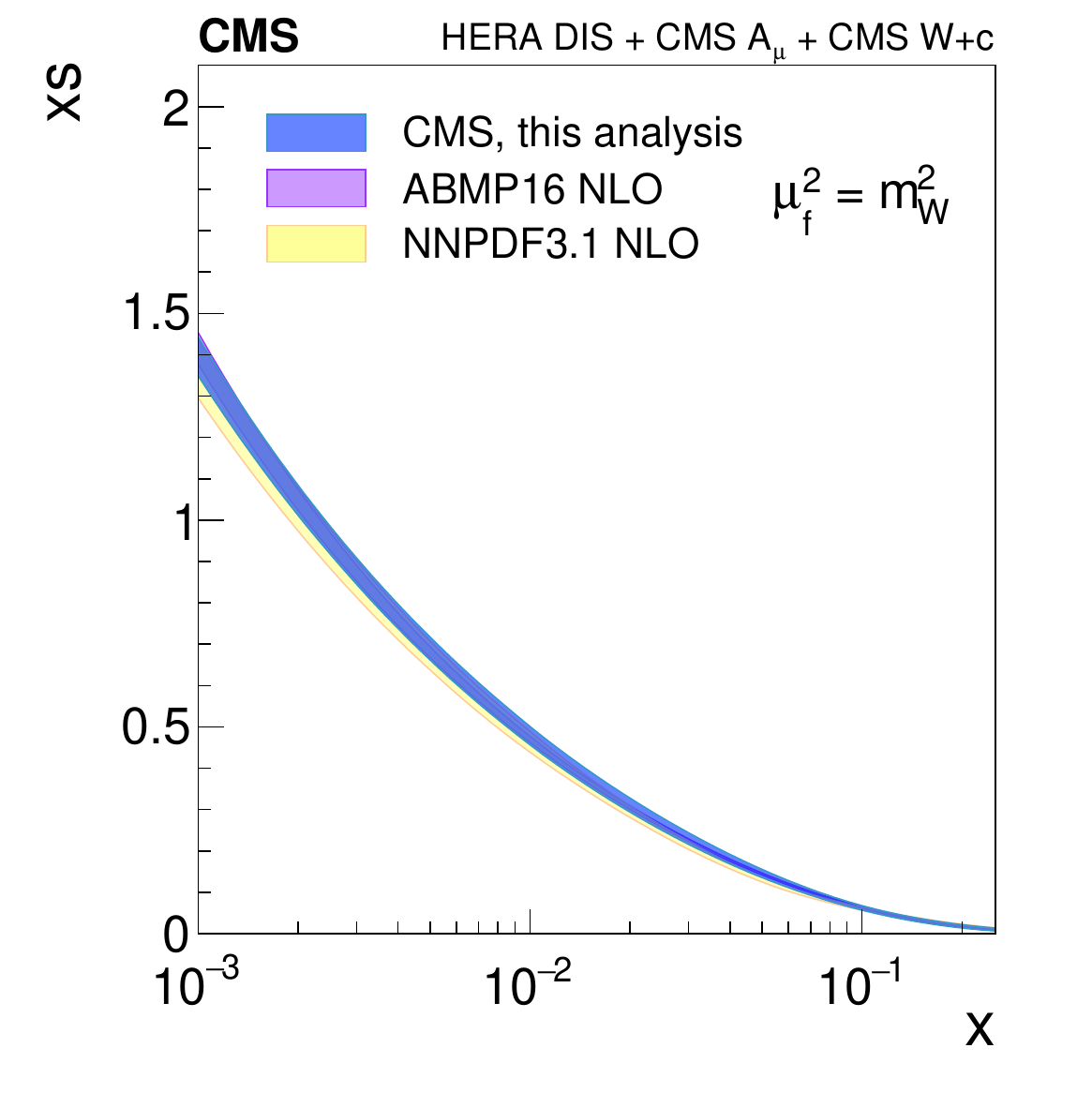}
  \includegraphics[width=0.48\textwidth]{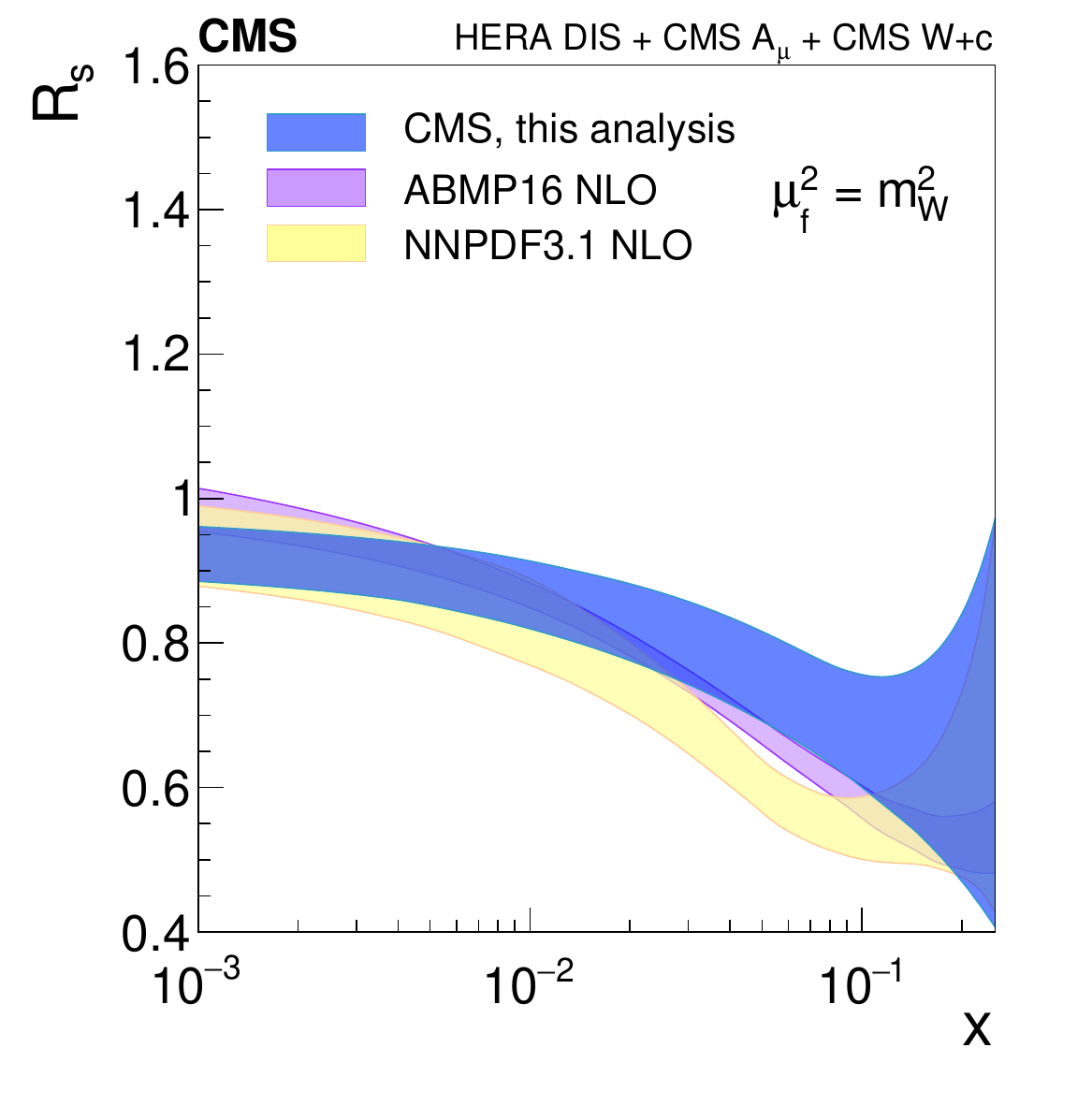}
  \includegraphics[width=0.48\textwidth]{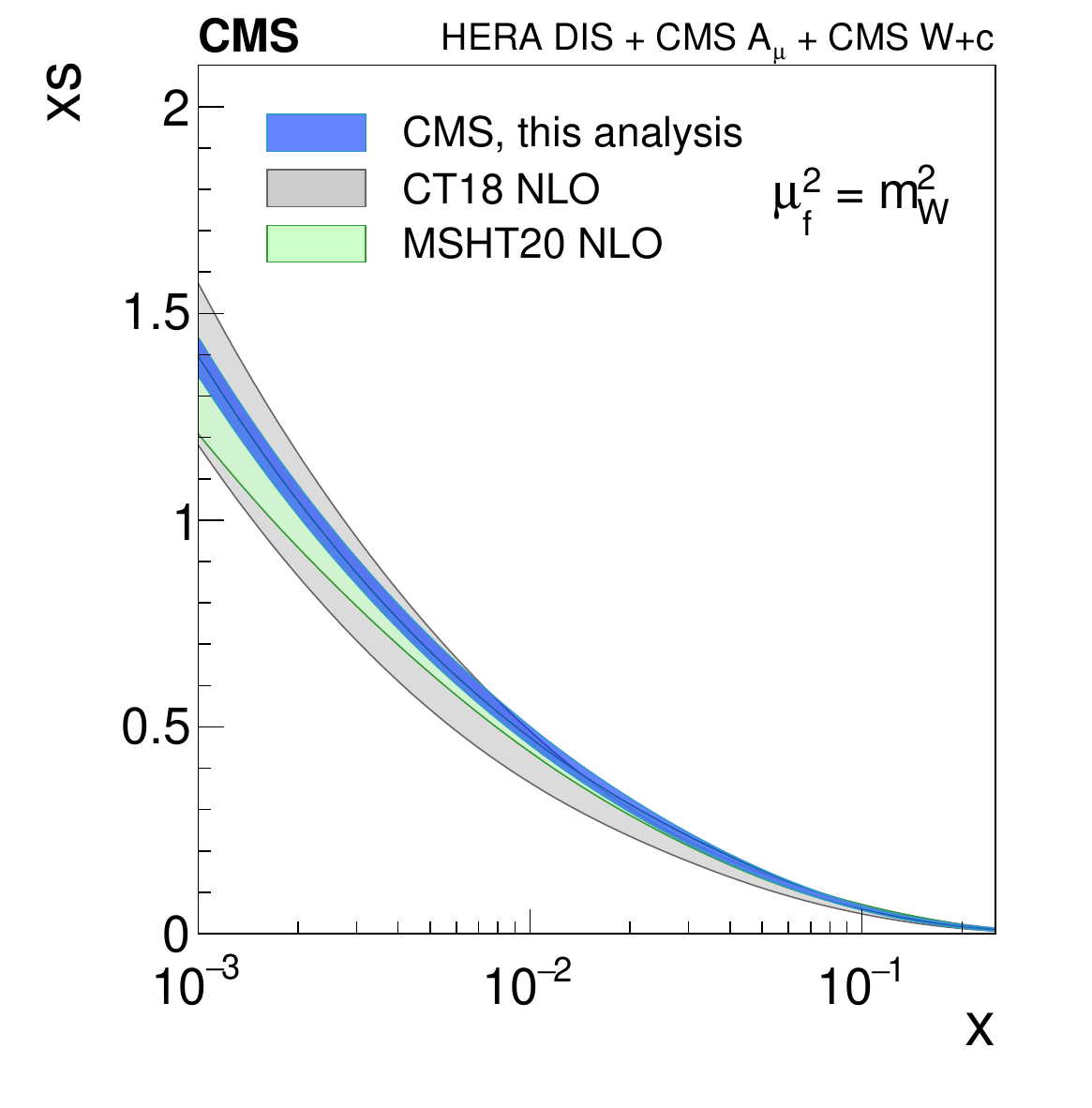}
  \includegraphics[width=0.48\textwidth]{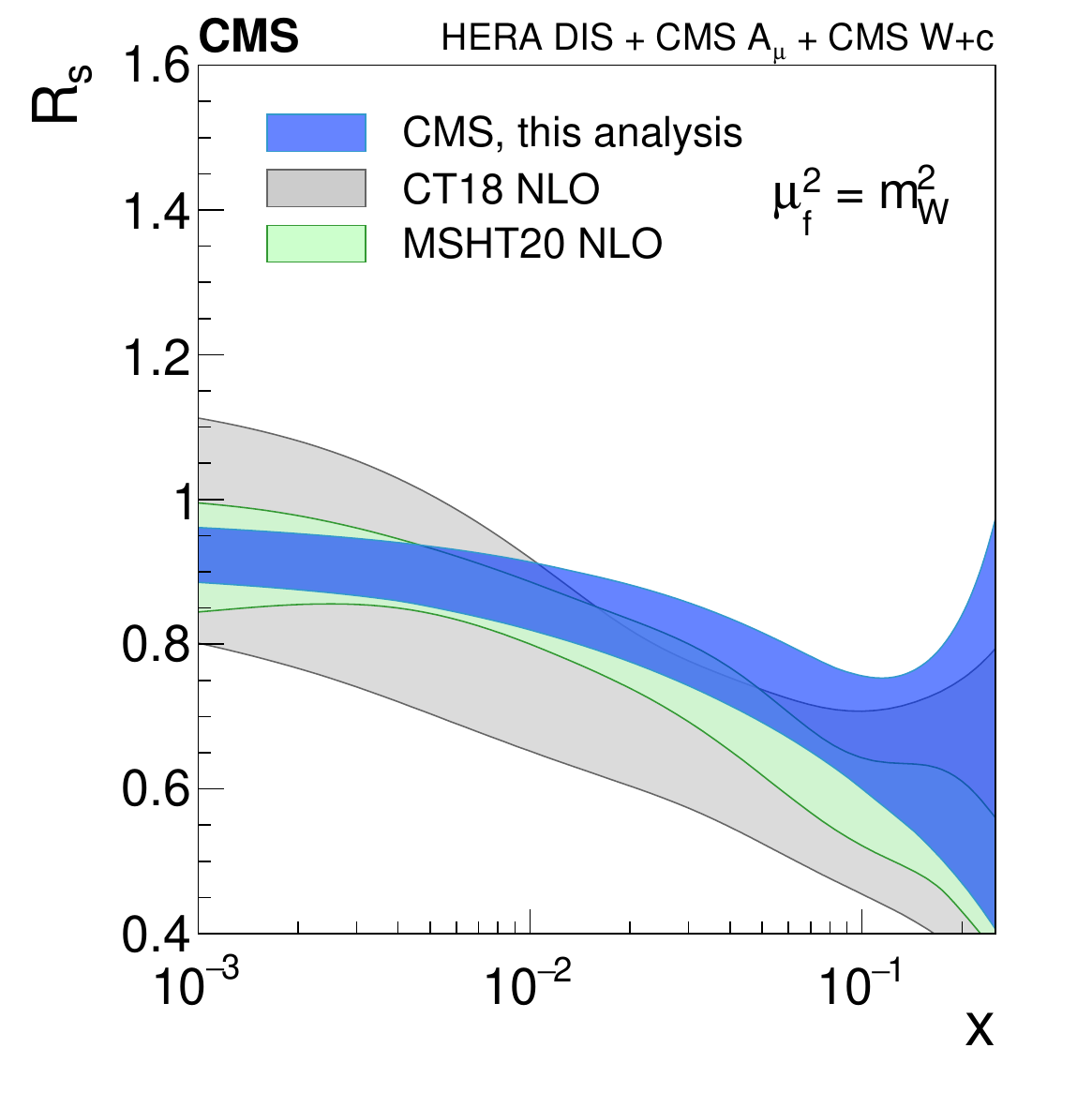}
  \caption{The strange quark distribution (left) and the strangeness suppression factor (right) as a function 
  of $x$ at the factorization scale of $m^2_{\PW}$. The results of the current analysis are shown together with those from 
  the global NLO PDFs, ABMP16 and NNPDF3.1 in the upper plot, and CT18 and MSHT20 in the lower one.
  This QCD analysis uses as input the combination of the inclusive deep inelastic scattering (DIS) 
  cross sections~\cite{Abramowicz:2015mha}, the CMS measurements of the lepton charge asymmetry in {\PW} boson production at 
  $\rts=7$ and 8\TeV~\cite{Chatrchyan:2013mza, Khachatryan:2016pev}, and the CMS measurements of $\Wc$ production at $\rts=7$~\cite{CMS-PAPER-SMP-12-002},
  ~8 (this analysis) and 13\TeV~\cite{CMS-PAPER-SMP-17-014}.  
}
  \label{s_rs_pdfs}
 \end{figure*}
\end{linenomath*}
\ifthenelse{\boolean{cms@external}}
{
}
{
\clearpage
}
\section{Summary}\label{sec:summary}

The associated production of a $\PW$ boson with a charm quark ($\Wc$) in proton-proton ($\Pp\Pp$) collisions at a centre-of-mass energy of 8\TeV is studied with a data sample collected by the CMS experiment corresponding to an integrated luminosity of 19.7\fbinv.
The $\Wc$ process is selected based on the presence of a high transverse momentum lepton (electron or muon) coming from a $\PW$ boson decay and a charm hadron 
decay.
Charm hadron decays are identified either by the presence of a muon inside a jet or by reconstructing a secondary decay vertex within a jet.
Inclusive and differential fiducial cross section measurements are performed with four different data samples 
(electron and muon $\PW$ boson decay channels and reconstruction of semileptonic and inclusive decays of charm hadrons). 
Cross section measurements are unfolded to the parton level.
The ratio of the cross sections of $\PWpc$ and $\PWmc$ is also measured. 
The results from the four different channels are consistent and are combined.

The measured fiducial $\Wc$ production cross section and the $(\PWpc)/(\PWmc)$ cross section ratio are:
\begin{linenomath*}
\ifthenelse{\boolean{cms@external}}
{
\begin{multline*}
\sigma(\ppWc) \, \mathcal{B}(\PWln)  = \\ 117.4 \pm 0.6 \stat \pm 5.6 \syst \unit{pb},
\end{multline*}
\begin{multline*}
\frac{\sigma(\ppWpc)}{\sigma(\ppWmc)}  = \\ 0.983 \pm 0.010\stat \pm 0.017 \syst. 
\end{multline*}
}
{
\begin{align*}
  \sigma(\ppWc) \, \mathcal{B}(\PWln) & = 117.4 \pm 0.6 \stat \pm 5.6 \syst \unit{pb},\\
  \frac{\sigma(\ppWpc)}{\sigma(\ppWmc)} & = 0.983 \pm 0.010\stat \pm 0.017 \syst. 
\end{align*}
}
\end{linenomath*}

The measurements are compared with the predictions of the \MADGRAPH MC simulation normalized to the NNLO cross
section prediction of inclusive $\PW$ production from \FEWZ. They are consistent within uncertainties.

The measurements are also compared with analytical NLO calculations from the \MCFM program using different NLO PDF sets. 
A fair agreement is seen in the differential cross section as a function of the absolute value of the pseudorapidity of the lepton from the $\PW$ boson.
Differences of ${\sim}10\%$ occur in the differential cross section as a function of the transverse momentum of the lepton in the 30--50\GeV range.

The combined measurement of the \Wc~ production cross section 
as a function of the absolute value of the pseudorapidity of the lepton
from the \PW boson decay is used in a QCD analysis at NLO, together with
inclusive deep inelastic scattering measurements from HERA and earlier results from CMS on \Wc~ production
and the lepton charge asymmetry in \PW boson production.
The strange quark distribution $x\PQs(x,\PGm_f^2)$ and the strangeness suppression
factor $R_{\PQs}(x,\PGm_f^2)$ = $(\PQs+\PAQs)/(\PAQu+\PAQd)$
are determined and agree with other NLO PDF sets such as ABMP16~\cite{ABMP16nlo},
NNPDF3.1~\cite{NNPDF31nlo},
CT18~\cite{CT18nlo},
and MSHT20~\cite{MSHT20nlo}.
The inclusion of the present results further constrains the strange quark distribution and the strangeness suppression factor.

\begin{acknowledgments}
  We congratulate our colleagues in the CERN accelerator departments for the excellent performance of the LHC and thank the technical and administrative staffs at CERN and at other CMS institutes for their contributions to the success of the CMS effort. In addition, we gratefully acknowledge the computing centres and personnel of the Worldwide LHC Computing Grid and other centres for delivering so effectively the computing infrastructure essential to our analyses. Finally, we acknowledge the enduring support for the construction and operation of the LHC, the CMS detector, and the supporting computing infrastructure provided by the following funding agencies: BMBWF and FWF (Austria); FNRS and FWO (Belgium); CNPq, CAPES, FAPERJ, FAPERGS, and FAPESP (Brazil); MES and BNSF (Bulgaria); CERN; CAS, MoST, and NSFC (China); MINCIENCIAS (Colombia); MSES and CSF (Croatia); RIF (Cyprus); SENESCYT (Ecuador); MoER, ERC PUT and ERDF (Estonia); Academy of Finland, MEC, and HIP (Finland); CEA and CNRS/IN2P3 (France); BMBF, DFG, and HGF (Germany); GSRI (Greece); NKFIA (Hungary); DAE and DST (India); IPM (Iran); SFI (Ireland); INFN (Italy); MSIP and NRF (Republic of Korea); MES (Latvia); LAS (Lithuania); MOE and UM (Malaysia); BUAP, CINVESTAV, CONACYT, LNS, SEP, and UASLP-FAI (Mexico); MOS (Montenegro); MBIE (New Zealand); PAEC (Pakistan); MSHE and NSC (Poland); FCT (Portugal); JINR (Dubna); MON, RosAtom, RAS, RFBR, and NRC KI (Russia); MESTD (Serbia); MCIN/AEI (Spain); MOSTR (Sri Lanka); Swiss Funding Agencies (Switzerland); MST (Taipei); ThEPCenter, IPST, STAR, and NSTDA (Thailand); TUBITAK and TAEK (Turkey); NASU (Ukraine); STFC (United Kingdom); DOE and NSF (USA).
  
  \hyphenation{Rachada-pisek} Individuals have received support from the Marie-Curie programme and the European Research Council and Horizon 2020 Grant, contract Nos.\ 675440, 724704, 752730, 758316, 765710, 824093, 884104, and COST Action CA16108 (European Union); the Leventis Foundation; the Alfred P.\ Sloan Foundation; the Alexander von Humboldt Foundation; the Belgian Federal Science Policy Office; the Fonds pour la Formation \`a la Recherche dans l'Industrie et dans l'Agriculture (FRIA-Belgium); the Agentschap voor Innovatie door Wetenschap en Technologie (IWT-Belgium); the F.R.S.-FNRS and FWO (Belgium) under the ``Excellence of Science -- EOS" -- be.h project n.\ 30820817; the Beijing Municipal Science \& Technology Commission, No. Z191100007219010; the Ministry of Education, Youth and Sports (MEYS) of the Czech Republic; the Deutsche Forschungsgemeinschaft (DFG), under Germany's Excellence Strategy -- EXC 2121 ``Quantum Universe" -- 390833306, and under project number 400140256 - GRK2497; the Lend\"ulet (``Momentum") Programme and the J\'anos Bolyai Research Scholarship of the Hungarian Academy of Sciences, the New National Excellence Program \'UNKP, the NKFIA research grants 123842, 123959, 124845, 124850, 125105, 128713, 128786, and 129058 (Hungary); the Council of Science and Industrial Research, India; the Latvian Council of Science; the Ministry of Science and Higher Education and the National Science Center, contracts Opus 2014/15/B/ST2/03998 and 2015/19/B/ST2/02861 (Poland); the Funda\c{c}\~ao para a Ci\^encia e a Tecnologia, grant CEECIND/01334/2018 (Portugal); the National Priorities Research Program by Qatar National Research Fund; the Ministry of Science and Higher Education, projects no. 14.W03.31.0026 and no. FSWW-2020-0008, and the Russian Foundation for Basic Research, project No.19-42-703014 (Russia); MCIN/AEI/10.13039/501100011033, ERDF ``a way of making Europe", and the Programa Estatal de Fomento de la Investigaci{\'o}n Cient{\'i}fica y T{\'e}cnica de Excelencia Mar\'{\i}a de Maeztu, grant MDM-2017-0765 (Spain); the Stavros Niarchos Foundation (Greece); the Rachadapisek Sompot Fund for Postdoctoral Fellowship, Chulalongkorn University and the Chulalongkorn Academic into Its 2nd Century Project Advancement Project (Thailand); the Kavli Foundation; the Nvidia Corporation; the SuperMicro Corporation; the Welch Foundation, contract C-1845; and the Weston Havens Foundation (USA).
\end{acknowledgments}

\bibliography{auto_generated}
\cleardoublepage \appendix\section{The CMS Collaboration \label{app:collab}}\begin{sloppypar}\hyphenpenalty=5000\widowpenalty=500\clubpenalty=5000\cmsinstitute{Yerevan~Physics~Institute, Yerevan, Armenia}
A.~Tumasyan
\cmsinstitute{Institut~f\"{u}r~Hochenergiephysik, Vienna, Austria}
W.~Adam\cmsorcid{0000-0001-9099-4341}, T.~Bergauer\cmsorcid{0000-0002-5786-0293}, M.~Dragicevic\cmsorcid{0000-0003-1967-6783}, J.~Er\"{o}, A.~Escalante~Del~Valle\cmsorcid{0000-0002-9702-6359}, R.~Fr\"{u}hwirth\cmsAuthorMark{1}, M.~Jeitler\cmsAuthorMark{1}\cmsorcid{0000-0002-5141-9560}, N.~Krammer, L.~Lechner\cmsorcid{0000-0002-3065-1141}, D.~Liko, T.~Madlener\cmsorcid{0000-0002-0128-6536}, I.~Mikulec, F.M.~Pitters, N.~Rad, J.~Schieck\cmsAuthorMark{1}\cmsorcid{0000-0002-1058-8093}, R.~Sch\"{o}fbeck\cmsorcid{0000-0002-2332-8784}, M.~Spanring\cmsorcid{0000-0001-6328-7887}, S.~Templ\cmsorcid{0000-0003-3137-5692}, W.~Waltenberger\cmsorcid{0000-0002-6215-7228}, C.-E.~Wulz\cmsAuthorMark{1}\cmsorcid{0000-0001-9226-5812}, M.~Zarucki\cmsorcid{0000-0003-1510-5772}
% ...
\vskip\cmsinstskip
% ...
1:~Also at TU~Wien, Wien, Austria\\
% ...
\end{sloppypar}
\end{document}